\documentclass{article}
\usepackage{amssymb}

\input{tcilatex}

\begin{document}

\title{Little groups of\ irreps of O(3), SO(3), and the infinite axial
subgroups}
\author{M\ J Linehan \\
School of Physics, University of Exeter, Exeter EX4 4QL, U.K.\\
and \and G E Stedman$^{\dagger }$ \\
Department of Physics and Astronomy, University of Canterbury, \\
Private Bag 4800, Christchurch 8020 New Zealand.}
\maketitle

\begin{abstract}
Little groups are enumerated for the irreps and their components in any
basis of O(3) and SO(3) up to rank 9, and for all irreps of C$_{\infty }$, C$%
_{\infty h}$, C$_{\infty v}$, D$_{\infty }$ and D$_{\infty h}$. The results
are obtained by a new chain criterion, which distinguishes massive
(rotationally inequivalent)\ irrep basis functions and allows for multiple
branching paths, and are verified by inspection. These results are relevant
to the determination of the symmetry of a material from its linear and
nonlinear optical properties and to the choices of order parameters for
symmetry breaking in liquid crystals.
\end{abstract}

\bigskip \bigskip

\bigskip

\footnote{%
E-mail: g.stedman@phys.canterbury.ac.nz}

\newpage

\section{Introduction}

The little group, or isotropy group, H$_{\lambda l}$ of a component $%
|\lambda l\rangle $ (basis function)\ of an irrep (irreducible
representation)\ $\lambda $ of a group G is the maximal symmetry group of
this component. That is, it is the maximal subgroup of G\ under each of
whose elements this component is invariant. It is not necessary that the
axis choices for G and H be identical. We define the little groups of $%
\lambda \left( \text{G}\right) $ as the set H$_{\lambda }=\left\{ \text{H}%
_{\lambda l}|l\in \lambda \right\} $ of all possible symmetry groups of any
single function or linear combination within this irreducible space. The
inclusion of all linear combinations ensures that this set is independent of
the choice of basis $\left\{ l\right\} $, making it relevant for all
physical applications.

Applications of the little group concept cover the gamut of theoretical
physics. In the group theoretical approach to symmetry breaking problems the
symmetry of the solutions does not depend on the details of the governing
equations. \ Complete solutions will still require analysis of the governing
equations, but the group theoretical selection rules may reduce the number
of possibilities to a manageable number.In recent years it has been
discussed in the physics literature in connection with internal symmetries
in a relativistic theory (e.g. Landsman and Wiedemann 1995, Kim 1996, 1997),
Lorentz-Gauge potentials (Leaf 1998), tachyons (Rembielinski 1997), the
Pauli-Lubanski spin vector (Nash 1997), finding the Higgs minima with the
aid of little groups of the irreps of the gauge group (Girardi \emph{et al.}
1982), space groups (e.g. Jaric 1983a, Jaric and Senechal 1984) and
aperiodic crystals (Konig and Mermin 1997).

The Landau theory is a cornerstone of our present understanding of second
order phase transitions. \ The reason the Landau theory is often
qualitatively accurate, despite major quantitative approximations, is that
it focuses on the exact symmetries involved in phase transitions. \ For
example, the transition may only be second order when the system symmetry
changes at the transition. \ Landau's seminal hypothesis is that the
thermodynamic potential of a system in the vicinity of a phase transition is
a functional of the symmetry-breaking part of the relevant system density,
namely the order parameter for the transition. Since the expansion should
provide a valid expression of the thermodynamic potential both above and
below the transition, it must have at least as high a symmetry as the more
symmetrical of the two phases. \ This is essentially an application of
Neumann's Principle. \ Since the transition occurs from the more symmetrical
symmetry group (say $G$) to a less symmetrical symmetry group (say $H$) one
may describe the system in terms of the irreps of $G$. \ The order parameter
belongs to the identity representation of $H$ but not to the identity
representation of $G$, and the little groups of the irreps of $G$ therefore
play an important role (Jaric 1986). \ A reduction of symmetry G $%
\rightarrow $ H $\subset $ G is most appropriately quantified by an order
parameter $Q$ which has H for its little group. If $Q$ corresponds to a
single irrep $\lambda $ of G, H is H$_{\lambda }$ as defined above. If we
can specify $Q$ as a particular basis function $|\lambda l\rangle $ in the
irrep space, H can be identified with H$_{\lambda l}$; however in practice
symmetry arguments alone are unlikely to give this further restriction.
Indeed the physics of the situation may rather require a reducible choice of 
$Q$, as discussed by Birman (1982). Successful applications include
superfluid He$^{3}$ (see Vollhardt and W\"{o}lfle 1990 for a comprehensive
review). \ In this case the symmetry breaking occurs from a seven parameter
Lie group that describes the spin angular momentum, orbital angular momentum
and gauge symmetry of the Cooper pairs that make up the superfluid. \ The
little group technique successfully predicts all known superfluid phases of
He$^{3}$ and provides the starting point for any phenomenological theory of
them. \ \ 

One interest of ours is in the optimal choice of symmetry of the order
parameter for each of various liquid crystal phase transitions. Because this
interest includes isotropic to nematic liquid crystals, the little groups
for irreps of O(3)\ are of interest, as are those of SO(3) for chiral
systems. D$_{\infty h}$ in non-chiral systems and D$_{\infty }$ in chiral
systems are similarly of interest for liquid crystal transitions from
smectic A to smectic C or to smectic B hexatic. Analyses of the symmetry of
order parameters in liquid crystal transitions have tended not to consider
explicitly the effects of the reduction in point group symmetry of the
orientational orderings, and have concentrated for example on such aspects
as the breakdown of translational symmetry in the density wave of smectic
liquid crystals. The little group concept has an important role in
optimizing the description of the point group symmetry reduction basic to
the onset of the various orientational orderings. This task requires a
knowledge of the little groups of irreps of O(3)\ (the symmetry of the
isotropic phase from which the nematic transition occurs) and of D$_{\infty
h}$ (the symmetry of the bond-orientational order in the nematic phase, to
be broken by the further transitions). The appropriate choice of order
parameter for each should reflect the nature of the symmetry being broken,
both by being written in terms of the irreps of the higher symmetry phase
and by having as little group the symmetry of the lower-symmetry phase.

Another application that we have in mind is to the determination of the
symmetry of a system by (for example) experimental determination of its
optical properties, both linear and nonlinear. It is known that the symmetry
of a system may be ascertained to some degree from its spectra, and that the
extent to which the symmetry can be pinned down varies with the technique
used.

There are several complications in the latter exercise. Many sets of tables
exist of the tensor components appropriate to (for example)\ phonon-assisted
Raman scattering (e.g. Loudon 1964) and second harmonic generation (e.g.
Giordmaine 1965, Yariv and Yeh 1984). Such tabulations note the nonzero and
equivalent elements. Based on a semiclassical approach, which has its own
subtle inadequacies (e.g. Andrews \emph{et al.}1998),\ these tables make
many implicit and simplifying assumptions, such as the use of a singlet
ground state. When the possibility of a degenerate ground state is
considered in Raman scattering, the optical properties of the system are
considerably more complicated (Churcher and Stedman 1981). Even the
absorption spectrum of a low-symmetry system in a general linear
polarisation cannot be written as a linear combination of the spectra for $%
x,y,z$ linear polarisation, because the spectrum is of second order and
requires cross terms in the amplitudes (Stedman 1985). Even so, these tables
indicate that one may discriminate between (for example)\ C$_{3h}$ and D$%
_{3h}$ systems through their differing tensorial form in second harmonic
generation.

However if the requirement is foregone that the axes of reference are fixed
and the question is asked whether two symmetries can be distinguished modulo
a relative rotation, then little group analysis proves that many structures
of different symmetry and with different property tensors behave
identically. One powerful application of these results is to property
tensors associated with certain optical effects (see for example Nye 1957).
It is useful for condensed matter or nonlinear optical experimenters to use
tabulations by point group of the symmetry conditions on ligand field
parameters (Newman 1971), also the scattering tensors for such processes as
phonon-assisted Raman scattering (Loudon 1964), second harmonic generation
and sum frequency generation (Giordmaine 1965)\ etc. In establishing the
non-zero and related elements of such tensors, it is normal to assume a
particular axis system. However many spectroscopic fitting techniques (such
as the fitting of the parameters in a crystal field Hamiltonian to an
observed energy level spectrum)\ are ignorant of the coordinate system used
in the theory, and are vulnerable to the associated introduction of
unphysically high numbers of parameters. Such results, familiar in the
folklore of crystal field theory, can be put on a firmer foundation by
recognising them as consequences of the identification of little groups for
the relevant O(3) tensors.

For example, in ligand field theory for a system of symmetry C$_{4h}$
inclusion of the parameter $A_{4-}^{4}$ (the coefficient of the
corresponding tesseral harmonic $Z_{m}^{l}$ with $l=4$, $m=4$-)\ in a level
fitting program leads to indeterminancy in the fitting procedure, because
its value is arbitrary in the sense that one may rotate the coordinates
about the $z$ axis by any angle $\theta $, changing the relative size of
this parameter and $A_{4+}^{4}$; hence a least squares fit becomes
indefinite through the ambiguity in the choice of $\theta $. One should
define $A_{4-}^{4}$ to be zero, as if the symmetry were D$_{4h}$. Section %
\ref{o3} gives a group-theoretic reason for this observation:\ D$_{4h}$ and
not C$_{4h}$ is a little group for $l=4^{+}$ (the parity of a crystal field
Hamiltonian is positive since it acts within $d$ electron manifold), and the
tensorial structure appropriate to D$_{4h}$ should therefore be used. The
relationship of this ambiguity to the $z$ rotation is illustrated in Section %
\ref{o3} As another example, systems of symmetry C$_{3h}$ and D$_{3h}$
cannot be distinguished through second harmonic generation measurements,
because the little groups appropriate to second harmonic generation
(corresponding to irreps 1$^{-}$, 2$^{-}$ and 3$^{-}$) include D$_{3h}$ and
not C$_{3h}$. Again the physical reason is that the extra parameters which
supposedly distinguish C$_{3h}$ from D$_{3h}$ can be removed by a suitable
(parameter-dependent) rotation of the system, in this case an $xy$ rotation
to make the coefficient of say $Z_{3-}^{3}$ vanish. As another example, the
dielectric tensor $\mathbf{\varepsilon }$ of a crystal usually has a higher
symmetry than the point group G\ suggests, being biaxial (diagonal) if G $%
\subseteq $ D$_{2h}$, uniaxial (diagonal and with $\varepsilon
_{xx}=\varepsilon _{yy}$) if G $\not\subseteq $ D$_{2h}$ but G $\subseteq $ D%
$_{\infty h},$and isotropic otherwise. The reason is conspicuous from
Section \ref{o3}; such a symmetric even-parity second rank tensor must
transform as $l^{\pi }=0^{+}+2^{+}$ under SO(3), and the relevant little
groups O(3), D$_{\infty h}$ and D$_{2h}$ dictate the optimal form of the
tensor, once a suitable rotation (in this case diagonalising the matrix)\ is
applied. It is not guaranteed that the axes of the dielectric tensor
coincide with any of the crystal axes. The same mathematics may be applied
to justify the manner in which the moment of inertia tensor of undergraduate
physics can be represented by an ellipsoid of revolution, even for objects
with far less (and even no) symmetry, when principal axes are chosen
suitably.

In the applied mathematics literature, two applications of the little group
concept are to buckling problems and to Rayleigh-B\'{e}nard instabilities in
thermal convection (Ihrig and Golubitsky 1984). Many problems in bifurcation
theory (e.g. Keller 1969) have the property that when a bifurcation occurs
in a stationary problem the symmetry of the solution is a subgroup of that
of the ground state. \ Consequently, this sort of problem often involves
finding symmetry breaking solutions to a system of non-linear partial
differential equations. \ The spherical B\'{e}nard problem forms an example
of such a bifurcation problem. \ It may involve the steady states in the
buckling of an elastic spherical shell (like a red blood cell) or convection
of a viscous fluid confined between two concentric spherical shells (like
the magma between the crust and the core of the earth). \ It has been
successfully approached via the Liapunov Schmidt reduction (Sattinger 1978),
which is contingent upon knowledge of the little groups of O(3) (Ihrig and
Golubitsky 1984). \ A similar situation is that of finding the equilibrium
of a rotating self-gravitating fluid. \ The relevant equations depend on the
square of the angular momentum. \ Consequently they and their unbifurcated
solutions (known as Maclaurin ellipsoids) are invariant under the group D$%
_{\infty h}$. \ The first symmetry breaking solutions were found by Jacobi
(1834) and are known as the Jacobi ellipsoids (symmetry group D$_{2h}$).
Poincar\'{e} (1885) made further contributions. \ It was not until
relatively recently (Constantinescu 1979) that complete (infinite) solutions
were found for bifurcations from both the Maclaurin and Jacobi ellipsoids
using the little groups of D$_{\infty h}$ and D$_{2h}$ respectively. \ 

The results presented here are therefore of immediate interest in the Landau
theory of liquid crystal phase transitions and in the identification of
molecular symmetries from their spectra. However they are of far more
general application, as indicated above.

\section{Chain criteria for little group identification\label{cc}}

Several authors have investigated the little groups of the rotation group
SO(3). No fully reliable algorithm for determining the little groups of O(3)
or SO(3)\ has previously been reported.

Listings have been made of the little groups of the irreps ($l^{\pi }$ where 
$l=0,1,2,...$ and $\pi =\pm 1$) of O(3). In many subsequent works, a chain
criterion has been the central consideration. Such criteria have been
discussed extensively (Birman 1966, Goldrich and Birman 1968, Cracknell 
\emph{et al.} 1976, Boccara 1973, Lorenc \emph{et al.} 1980, Birman 1982,
Jaric 1982, Przystawa 1982, Gaeta 1990). This chain criterion has passed
through several modifications in the literature, and the little group
listing alters with each modification. Problems in the analysis of Boccara
(1973), for example, are mentioned in Appendix 1.

\subsection{Michel criterion}

A systematic determination of the little groups of SO(3) was made by Michel
(1980). He used a form of the chain criterion (his lemma 2) except in the
case of all cyclic groups, where little groups were determined by
inspection. \ The present work gives two small corrections to Michel's list.
\ First, SO(3) is only trivially a little group for a non-identity irrep. \
Second, Michel's identification of D$_{2}$ as a little group for $l=3$ is
incorrect. This indicates an inadequacy in his chain criterion. The aim of
the present work is to find an approach that generates the correct little
groups and to extend the analysis to find the most general vector in the
group representation space for a given little group, generalising these
results to cover irreps of the full rotation group O(3). \ Michel notes that
this is trivial in the case of positive parity representations if one knows
the results for SO(3). \ This is because the full rotation group is an outer
direct product group; O$(3)=$SO$(3)\times $Z$_{2}=$Z$_{2}\times $SO$(3)$
where Z$_{2}$ denotes the parity (inversion) group. \ Thus, if a
representation belongs to the identity irrep of Z$_{2}$, its little groups
will simply be the direct product of Z$_{2}$ with the relevant SO(3) little
groups. \ In the case of negative parity no such simple prescription exists
as improper operations may still be present (as in the intrinsic symmetry of
a true vector such as the electric field, the symmetry of which is C$%
_{\infty v}$).

In his lemma 2 and Table AII Michel (1980) used a group subduction criterion
which may be rephrased in the following form. If H is to be a little group
of an orthogonal irrep $\lambda \left( \text{G}\right) $, for a strictly
larger intermediate subgroup H$^{\prime }$ such that H$\subset $H$^{\prime
}\subset $G, and in particular for a group H$^{\prime }$ adjacent to H in
this group coupling chain, then 
\begin{equation}
c_{\lambda }\left( \text{H}^{\prime }\right) <c_{\lambda }\left( \text{H}%
\right) .  \label{Michel}
\end{equation}%
In this equation $c_{\,\lambda }\left( \text{H}\right) $ is the subduction
frequency for H in $\lambda $(G), namely the number of occurrences of the
identity irrep $0\left( \text{H}\right) $ in $\lambda \left( \text{G}\right) 
$; it is readily determined from the Weyl trace formula $c_{\lambda }$(H) $%
=\sum_{h\in H}\chi ^{\lambda (G)}\left( h\right) /\left| \text{H}\right| $.

From the above chain criterion the largest group H that has a particular $%
c_{\lambda }($H$)$ is the little group of that particular linear combination
of basis functions. \ If one moves from a little group to the next group up
the group chain then the value of $c_{\lambda }($H$)$ decreases because that
increase in symmetry removes one or more of the basis functions that were
allowed at the lower symmetry. \ The little group is then the maximal
symmetry of some linear combination of basis functions. \ A simple
illustration of the application of this chain criterion is given in Appendix
1, where the tables give the group-subgroup branching relations needed for
the use of the chain criterion for the 32 crystallographic point groups and
also for other groups of interest in our applications. The entries are the
subduction frequencies, or number of times the invariant 0(H) occurs in $%
\lambda $(G). Equation (\ref{Michel}) is a useful necessity condition for H
to be a little group.

Jaric (1982) noted the failure of the chain criterion to be sufficient for a
Lie group. There are two reasons for the lack of sufficiency. Even for
finite groups equation (\ref{Michel}) can fail to be a sufficiency
condition, because multiple group-subgroup coupling chains connecting H$%
^{\prime }$ and H need special consideration. In addition, the little group
is the \emph{maximal }symmetry group its symmetry elements do not have to
share the axes of the same elements in the parent group. Every function in
the basis set can have its own orientation relative to the supergroup axes.
For example, C$_{s}$ is not a little group of 1$^{-}($O(3)) in spite its
conformity to this chain criterion. The conformity is shown because the
vector irrep 1$^{-}\left( \text{O}(3)\right) $ branches twice to 0$\left( 
\text{C}_{s}\right) $, once more than in any of its supergroups C$_{5h}$ $,$
C$_{5v}$, C$_{2h}$, C$_{2v}$, C$_{3v}$ which are maximally connected (i.e.
which have no intervening group in the chain). However these numbers reflect
the fact that all the components are tied to the same choice of rotation
axes and mirror plane. When the axis choice can be tailored individually to
the basis functions their symmetry is seen to be much higher than this, and
in fact the little group of the functions in 1$^{-}\left( \text{O}%
_{3}\right) $ is uniquely C$_{\infty v}$, the intrinsic symmetry group of
any polar vector. As another example, $2($SO$\left( 3\right) )$ reduces to
more invariants in the maximal subgroup C$_{2}$ of D$_{2}$ than it does in D$%
_{2}$ itself; again more invariants appear at C$_{1}$ level. However C$_{1}$
and C$_{2}$ are not little groups of $2($SO$\left( 3\right) )$. The (real
and orthogonal)\ basis functions $\ Z_{1+}^{2}=xz$, $Z_{2+}^{2}=x^{2}-y^{2}$%
\ in $2\left( \text{SO}\left( 3\right) \right) $ both have D$_{2}$\
symmetry, but about different axes. In addition, their linear combination
has no 2-fold symmetry about an axis in the same plane; yet it has D$_{2}$\
symmetry about oblique axes.

These examples show that the symmetries of all these functions cannot be
tested at once within a global choice of axes by counting the increases in
the number $c_{\lambda }$ of invariants as the symmetry is lowered simply by
removing group elements relative to those axes. An approach is needed in
which bases are allowed to be fully flexible so as to investigate the
maximal symmetry of any function.

In particular, Michel (1980) was incorrect to assign D$_{2}$ as a little
group of $l=3$ under SO(3) (see Appendix 1).

\subsection{Ihrig and Golubitsky criterion}

Michel (1980) left the Parthian challenge: ``We leave to the reader the
study of the representations 1-''. \ The challenge was taken up by Ihrig and
Golubitsky (1984). Their work is the most sophisticated to date and has been
accepted by subsequent workers as the definitive result.

Ihrig and Golubitsky (1984) observe: ``Unfortunately, Michel's criterion for
determining when a subgroup is actually an isotropy subgroup (lemma 2) is
incorrect as stated,'' \ and ``In section 5 we give a correct version of
this lemma. \ See lemma 5.3. Its proof is involved. It seems likely that the
condition we give is both necessary and sufficient though we have not been
able to prove this.'' \ Our results show that their lemma 5.3 is necessary
but not sufficient for the group O(3), and not all their identifications are
correct.

Initially, Ihrig and Golubitsky (1984) undertake to find the maximal little
groups of O(3), stating: ``it is harder for a system to break more
symmetries than less.'' \ This, the so-called maximality conjecture, has
been part of the folklore of phase transition and gauge field theory for
twenty years. Its general applicability was disproved by Jaric (1983b). \
Ihrig and Golubitsky (1984) then embark on the task of finding all of the
little groups of O(3), a task described as ``a much more difficult
calculation''. \ 

Ihrig and Golubitsky's (1984)\ central result, proposition 5.3, is another
refinement of the chain criterion. Rather than depending on an increase in
the total subduction frequency, Ihrig and Golubitsky propose that the
following inequality must be satisfied in a H$^{\prime }$ $\supset $ H if H
is to be a little group of $\lambda \left( \text{G}\right) $:

\begin{equation}
c_{\lambda }\text{(H}^{\prime }\text{)}-\dim N_{G}\left( \text{H}^{\prime
}\right) <c_{\lambda }\text{(H)}-\dim N_{\text{G}}\left( \text{H,H}^{\prime
}\right)  \label{IG1}
\end{equation}

In this the following definitions are used. \ As above, $c_{V}$(G) (which
Ihrig and Golubitsky write as $\dim V^{G}$)\ is the subduction frequency for
the group G in the vector $V$, namely the number of occurrences of the
identity irrep $0$(G) in the set $V$. In the case of $c_{\lambda }$ $%
V=V_{\lambda }$, the set of basis functions of the irrep $\lambda \left( 
\text{G}\right) $.

The normaliser $N_{G}\left( \text{H}\right) $ of any H $\subset $G is the
largest subgroup of $G$ that contains $H$ as an invariant subgroup:\ $%
N_{G}\left( \text{H}\right) \equiv \left\{ g\in \text{G }\mid \text{ }g\text{%
H}g^{-1}=\text{H}\right\} $ (see Fraleigh 1994) . \ For K $\subset $H, $%
N_{G}\left( \text{K,H}\right) \equiv \left\{ g\in \text{G }\mid \text{ }g%
\text{H}g^{-1}\supset \text{K}\right\} .$ For a group, $\dim $G is the Lie
group dimension of the manifold of G, namely the number of infinitesimal
generators in G. Because we deal only with subgroups of O(3), $\dim $G $=3$
if G $=$ O(3) or SO(3), $\dim $G $=1$ if G is one of the infinite axial
groups, and $\dim $G $=0$ if G is a finite group. The normaliser of a finite
subgroup of O(3) is usually also a finite group and so of dimension zero. \
The exceptions are the normalizers of any of the Abelian subgroups of O(3).
\ Any subgroup of an Abelian group is an invariant subgroup, so that the
normaliser of any Abelian subgroup of O(3) will always contain C$_{\infty }$
and will thus always be one-dimensional. \ 

This refinement of the chain criterion by Ihrig and Golubitsky (1984),
introducing the dimensions of the normalisers, shows cognisance of the basic
problems with the Michel criterion, and is closer to a sufficiency
condition. The differences are discussed in Sections \ref{mass}, \ref{o3-2}
and Appendix 1. We note here that the little groups of SO(3) generated by
their method mostly agree with Michel (1980), including the erroneous
assignment of D$_{2}$ in $l=3$; they disagree with Michel (1980)\ by
assigning O and not T\ as a little group of $l=3$; these changes are
retrograde as erroneous. For positive parity irreps of O(3), their results
are completely incorrect; none of the groups they list are little groups.
For negative parity irreps in O(3)\ their results are not consistently
presented. On the most favourable rationalisation of this problem, they are
incorrect in assigning T as a little group in $l=3^{-},4^{-}$. Hence
equation (\ref{IG1}) is not sufficient.

\subsection{Massive chain criterion \label{mass}}

We introduce now a modified criterion which we call the massive chain
criterion; we find it to be fully reliable as a sufficiency as well as
necessity condition within subgroups of G = O(3). This criterion states that
H $\subset $ G is a little group of an orthogonal irrep $\lambda \left( 
\text{G}\right) $ if and only if for each strictly larger and adjacent\
group H$^{\prime }$ (so that H $\subset $ H$^{\prime }\subset ...\subset $
G), \emph{\ } \emph{\ }

\begin{equation}
c_{\lambda }(\text{H}^{\prime })-f_{\lambda }^{0}\left( \text{H}^{\prime
}\right) <c_{\lambda }(\text{H})-f_{\lambda }^{0}\left( \text{H}\right) .
\label{mcc1}
\end{equation}

Here we call $f_{\lambda }^{0}\left( \text{H}\right) $ the massless
subduction frequency; it will be defined below. $f_{\lambda }^{0}\left( 
\text{H}\right) $ redeems the failure of earlier chain criteria to recognise
adequately that the shape of a basis function is independent of any O(3)
rotation, and that the attendant freedom of its axis choice has to be taken
into account in determining whether it is an invariant\thinspace\ with
respect to any potential little group H. Comparing equations (\ref{mcc1}), (%
\ref{Michel}) and (\ref{IG1}), we note that the massless subduction
frequency $f_{\lambda }^{0}\left( \text{H}\right) $ plays the role of Ihrig
and Golubitsky's correction term $\dim N_{\text{G}}\left( \text{H,H}^{\prime
}\right) $ in refining the Michel criterion. We shall illustrate that $%
f_{\lambda }^{0}$(H) can differ from $\dim N_{\text{G}}\left( \text{H,H}%
^{\prime }\right) $ by more than a constant even in subgroups of O(3).

Another practical difference with both Michel (1980) and with Ihrig and
Golubitsky (1984)\ is the emphasis on checking the criterion in each
possible coupling chain H$^{\prime }\supset $ H.\emph{\ }This is an
important check when there are multiple chains, to avoid H being credited
with little group character when that application should really be to some
higher group. This explains at least some of the explicit errors in earlier
work.

Let $V_{\lambda }\left( \text{H}\right) $ be the subset of \ functions in $%
V_{\lambda }$ (a set of orthogonal functions forming a basis of $\lambda
\left( \text{G}\right) $) that are invariant under H. The basis is chosen to
maximise the number of functions in the subset $V_{\lambda }\left( \text{H}%
\right) $. The dimension of $V_{\lambda }$ is therefore $c_{\lambda }$, the
number of occurrences of 0$\left( \text{H}\right) $ in $\lambda $. At this
stage no transformations of G may be applied to demonstrate such an
H-invariance; the basis functions, once chosen, are fixed and elements of G
that are retained in its subgroup H have unchanged orientations.

We now define the massless subduction frequency by determining the extent to
which the members of $V_{\lambda }($H) are in fact equivalent under
transformations of G. We partition $V_{\lambda }\left( \text{H}\right) $
into two sets, a `massless' subset $V_{\lambda }^{0}\left( \text{H}\right) $
and a `massive' subset $V_{\lambda }^{m}\left( \text{H}\right) $, so that
each member of $V_{\lambda }^{0}\left( \text{H}\right) $ when transformed by
a suitable element of G is identical to some member of $V_{\lambda
}^{m}\left( \text{H}\right) $. The partition is made so as to maximise the
dimension $f_{\lambda }^{0}\left( \text{H}\right) $ of $V_{\lambda
}^{0}\left( \text{H}\right) $ within the irrep $\lambda $. In other words,
G-equivalent functions in $V_{\lambda }$ are separated, one being
partitioned off to $V_{\lambda }^{0}\left( \text{H}\right) $ until no two
members of $V_{\lambda }^{m}\left( \text{H}\right) $ are identical under a
transformation of G. $V_{\lambda }^{m}\left( \text{H}\right) $ contains one
and only one of each of the shapes represented by the members of $V_{\lambda
}$; $V_{\lambda }^{0}\left( \text{H}\right) $ contains any duplicate shapes
(functions which are equivalent under G transformation). This defines the
`massless subduction frequency' $f_{\lambda }^{0}\left( \text{H}\right) $.
The massive subduction frequency is the dimension $f_{\lambda }^{m}\left( 
\text{H}\right) $ of $V_{\lambda }^{m}\left( \text{H}\right) $, and is
therefore given by\qquad \qquad 
\begin{equation}
f_{\lambda }^{m}\left( \text{H}\right) =c_{\lambda }-f_{\lambda }^{0}\left( 
\text{H}\right) .  \label{massdef}
\end{equation}

The terms `massless' and `massive' are chosen in analogy with Higgs theory
(see for example Weinberg 1996) where a unitarity transformation (the
analogue of the G operations above, and in particular to rotations for
SO(3))\ removes the basis functions of the gauge group that correspond to
massless particles. The analogy is particularly close when little groups are
applied to spontaneous symmetry breaking. Only massive components contribute
to the homogeneous Hamiltonian. \ The massless components become important
in the inhomogeneous part of the Hamiltonian because they give rise to
Goldstone modes, and the number of Goldstone modes must equal the number of
massless components.

With these definitions, equation (\ref{mcc1}) then gives%
\begin{equation}
f_{\lambda }^{m}\left( \text{H}^{\prime }\right) <f_{\lambda }^{m}\left( 
\text{H}\right) .  \label{mcc2}
\end{equation}%
When this inequality holds, a new shape (characterised independently of its
orientation)\ has appeared, and because it is invariant under H\ and not
under any supergroup H$^{\prime }$ its symmetry is H. This is the condition
for H to be a little group of $\lambda \left( \text{H}\right) $. By
construction, then, equation (\ref{mcc1}) takes fully into account the
complications in the subduction formula caused by the freedom of axis choice.

We now search for an algorithm to calculate the massless subduction
frequency $f_{\lambda }^{0}\left( \text{H}\right) $ in subgroups of G = O(3)
and SO(3). First, the value of $f_{\lambda }^{0}\left( \text{H}\right) $ is
limited by the need to have at least one massive function, $f_{\lambda
}^{m}\left( \text{H}\right) \geq 1$ within a finite subspace $V_{\lambda }$
of dimension $\dim V_{\lambda }$ (or $\left| \lambda \right| $). Hence $%
f_{\lambda }^{0}\left( \text{H}\right) \leq \dim V_{\lambda }-1$, which is $%
2l$ in the irrep $l^{\pi }$(O(3)). The next aspect is the extent to which
linearly independent functions can be generated from any basis function
while respecting H symmetries. The general rule for this pays particular
attention to the degrees of freedom of the group; linear independence is
then possible because of the continuous value of the possible rotation
(Ihrig and Golubitsky 1984). There are three cases to consider. We summarise
the results here; illustrations of such points are discussed in the next
section.

First, if there are no symmetry axes in H (i.e. for H = C$_{i}$ or C$_{1}$),
all basis functions are H-invariant; $V_{\lambda }\left( \text{H}\right)
=V_{\lambda }$; any rotation of O(3)\ will carry any basis function into
another H-invariant function. Using the three generators of SO(3), 3
linearly independent functions can be generated from any member of $%
V_{\lambda }^{m}\left( \text{H}\right) $ by rotation if the space is
sufficiently big ($\dim V_{\lambda }>2)$, so that in these cases $f_{\lambda
}^{0}\left( \text{H}\right) =3.$

Second, for the groups H = O(3), SO(3), D$_{\infty h}$, D$_{\infty }$, C$%
_{\infty v}$, Y$_{h}$, Y, O$_{h}$, O, T$_{h}$, T$_{d}$, T, D$_{nh}$, D$_{nd}$%
, D$_{n}$, C$_{nv}$ there are at least two non-collinear symmetry axes.
Hence any member of $V_{\lambda }\left( \text{H}\right) $, being invariant
under H, must also have a fixed orientation, and must be uniquely aligned
with these two axes. No continuous rotational degrees of freedom exist to
preserve the H-invariance and yet secure linear independence of any basis
function, and so no massless basis functions are possible: $f_{\lambda
}^{0}\left( \text{H}\right) =0$.

Third, if all the symmetry axes of H are the same, as for H = C$_{\infty h}$%
, C$_{\infty }$, C$_{nh}$, C$_{ni}$, C$_{n}$, S$_{n}$, C$_{s}$ at most one
generator of O(3)\ can act to change the orientation of any function
nontrivially. In the case of the abelian groups H = C$_{\infty h}$, C$%
_{\infty }$ $\dim V_{\lambda }=1$, there is no possibility of a massless
function and even this rotation is powerless to make the original basis
function $\exp im\phi $ linearly independent of its rotated form; $%
f_{\lambda }^{0}\left( \text{H}\right) =0$ in these groups. \ In the
remaining groups (C$_{nh}$, C$_{ni}$, C$_{n}$, S$_{n}$, C$_{s}$) no such
simplification arises and, corresponding to the number of generators, $%
f_{\lambda }^{0}\left( \text{H}\right) =1$ if $\dim V_{\lambda }>1.$

We now enshrine this argument and its conclusions in a formula. The number
of degrees of freedom of the group H, the Lie dimension of the normaliser
group of H, is defined as $\dim N_{G}\left( \text{H}\right) $; it\ gives
information on the number of independent H-invariant basis functions which
might be obtained from a member of the massive subset by an equal number of
types of rotations (Ihrig and Golubitsky 1984). Because of the
above-mentioned problem with C$_{\infty }$ and C$_{\infty h}$ (and this is
one point of departure from Ihrig and Golubitsky\ 1984\ in this paper), we
should subtract from $\dim N_{G}\left( \text{H}\right) $ the Lie dimension
of H, $\dim $H$;$ this connotes the one situation in which one degree of
freedom fails to generate an independent basis function (the third case
above). The fact that this is the only category of exceptional cases is
confirmed by inspection, and that confirmation (together with the provision
of a number of illustrations)\ is the role of Sections \ref{doo}, \ref%
{Strategy} and \ref{o3} in the argument of this paper. Taking into account
the dimensional restriction discussed above, we may then define%
\begin{equation}
\bar{f}_{\lambda }\left( \text{H}\right) =\dim N_{G}\left( \text{H}\right)
-\dim \text{H, \thinspace \thinspace \thinspace }f_{\lambda }^{0}\left( 
\text{H}\right) =\min \left[ \left( \dim V_{\lambda }-1\right) ,\,\,\bar{f}%
_{\lambda }\left( \text{H}\right) \right] .  \label{massless}
\end{equation}%
This completes the definition of the massive chain criterion. The results of
its application are tabulated in Appendix 2.

Note that $N_{G}\left( \text{H}\right) \supseteq $ H, so that $\dim
N_{G}\left( \text{H}\right) -\dim $H $\geq 0$. The number of massless
components of an irrep that subduces some group is not just the number of
degrees of freedom in orienting the basis functions, but this quantity minus
the number of generators in the group. The term $\left( \dim V_{\lambda
}-1\right) $ is important only for small irreps, with $\left| \lambda
\right| \leq 3$, which are easily dealt with by inspection.

For $\left| \lambda \right| >3$, the Ihrig and Golubitsky (1984)\ criterion,
equation (\ref{IG1}), and the massive chain criterion of equation (\ref{mcc1}%
) agree if

\begin{equation}
\dim N_{\text{G}}\left( \text{H,H}^{\prime }\right) =\dim N_{G}\left( \text{H%
}\right) +\dim \text{H}^{\prime }-\dim \text{H.}  \label{xxx}
\end{equation}%
In many cases this holds true. \ Ihrig and Golubitsky consider the case of H 
$=$ C$_{n}$ and H$^{\prime }=$ C$_{\infty }$ in some detail and find that $%
\dim N_{\text{G}}\left( \text{H,H}^{\prime }\right) $ $=$ $\dim N_{G}\left( 
\text{H}\right) .$ Our above-mentioned point of departure reflects the fact
that $\dim $C$_{\infty }=1$ although $\dim $C$_{n}=0$, violating equation (%
\ref{xxx}) and making the increase in the number of massless components in
going from C$_{\infty }$ to C$_{n}$ equal to one, not zero. This illustrates
the insufficiency of the criterion of equation (\ref{IG1}). \ The difference
between $f_{\lambda }^{0}\left( \text{C}_{\infty }\right) $ and $f_{\lambda
}^{0}\left( \text{C}_{n}\right) $ explains why we use a stronger criterion
than do Ihrig and Golubitsky (1984). \ The above analysis certainly confirms
that Ihrig and Golubitsky's proposition 5.3 is a necessity condition within
the group O(3), and as such is a stronger condition than that of Michel
(1980). \ 

Although the massive chain criterion only provides a slightly stronger
inequality than that of Ihrig and Golubitsky's work for some cases in
respect of O(3), it is unwise to conclude that these two different
inequalities will provide much the same results when applied to more
complicated Lie groups. \ Indeed, the fact that these two different
inequalities provide similar results for O(3) can be understood through
consideration of the concept of the normaliser group. \ Since the normaliser
of any group with respect to O(3) is by definition a subgroup of O(3), it
follows that its dimension can only be three (if it is O(3) or SO(3)), one
(if it is one of the infinite axial groups) or zero (if it is any finite
group). \ The dimension of a Lie group is identified in all of this work
with the dimension of its group manifold. \ If one considered a group with a
larger manifold (for example the group SU(5) which is 24 dimensional) then
it seems likely that discrepancies could become much more significant.

This analysis explains why, when inequality 5.3 of Ihrig and Golubitsky
(1984) is so similar to equation (\ref{mcc2}), these authors identify some
little groups incorrectly. \ The case of the group D$_{2}$ for $l=3($SO$%
(3))\;$ is instructive; this case is also incorrectly identified as a little
group by Michel (1980). \ Previous authors would not have found this to be a
little group if they had deployed the correct form of the chain criterion
for dealing with multiple inequivalent group chains, namely, if they had
required the criterion to hold for each possible H$^{\prime }$ in H$^{\prime
}\supset $ H. \ 

Finally, we note that Ihrig and Golubitsky's (1984)\ results (corollaries
6.7, 6.9) and our results (Section \ref{gen}) for little groups of O(3) for $%
l\geqslant 30$ are in complete agreement.

\section{The infinite axial groups\label{doo}}

We determine now the little groups of all irreps of C$_{\infty }$, C$%
_{\infty h}$, C$_{\infty v}$, D$_{\infty }$ and D$_{\infty h}$. While some
of these are well known, this is a complete list and the results give a
useful illustration of the concept of massless functions, a vital step to
the recognition of a more general chain criterion. One example of the
physical interest of these groups is that in nematic liquid crystals we
start not from O(3) but from D$_{\infty }$ or D$_{\infty h}$, since it is
the reduction of this symmetry which may be the decisive step.

The group C$_{\infty }$, isomorphic to the group $U(1)$, has one-dimensional
irreps; for the irrep labelled $m$ the basis function is the complex number
of unit modulus: $m\rightarrow e^{im\phi }$. Inspection or Michel's chain
criterion (equation \ref{Michel}) show that the little groups are $C_{\infty
}$ for $m=0$ and $C_{\left| m\right| }$ for all other $m$.\ 

In C$_{\infty h}$ also the irreps are one-dimensional. To apply the chain
criterion we need the branchings C$_{\infty h}\supset $C$_{nh}$, C$_{ni}$ ($%
n $ odd) , C$_{\infty }$; C$_{nh}$ ($n$ even)$\supset $C$_{n},$ C$_{i}$, C$%
_{s} $; C$_{nh}$ ($n$ odd)$\supset $C$_{n},$ C$_{s}$; C$_{ni}\supset $C$%
_{n}, $ C$_{i};$C$_{n}$ $\supset $C$_{1}$ the lower-order cases of which are
tabulated in Appendix 2. \emph{\ } The little groups (with the relevant
irreps in brackets)\ are as follows: C$_{\infty h}\left( 0^{+}\right) $, C$%
_{\infty }\left( 0^{-}\right) $, C$_{i}\left( 1^{+}\right) $, C$_{s}\left(
1^{-}\right) $; C$_{nh}\left( n^{+}\right) $ and C$_{n}\left( n^{-}\right) $
for $n$ even, $n\geq 2;$ C$_{ni}\left( n^{+}\right) $ and C$_{nh}\left(
n^{-}\right) $ for $n$ odd, $n\geq 3.$

The one-dimensional irreps of the groups $C_{\infty v},$ $D_{\infty }$ and $%
D_{\infty h}$ may be dealt with similarly, generalising the tabulated
branchings to C$_{\infty v}\supset C_{nv}$, C$_{\infty };$ C$_{nv}\supset $C$%
_{n}$, C$_{s}$. These groups all have two-dimensional representations, and
the Michel chain criterion is not sufficient. Consider the first two
dimensional irrep $1$ of C$_{\infty v}$. \ The only non-zero subduction
frequencies are $c_{1}(C_{s})=1$ and $c_{1}(C_{1})=2$. \ 

$C_{s}$ is clearly a little group for the irrep 1, because C$_{s}$\ is the
first group for which the massive subduction frequency is non-zero. If the
dimension of the representation vector is one, it cannot have any massless
components\textrm{. }\ Because there exists only a single axis of symmetry
(the direction perpendicular to the plane) and no infinitesimal generators,
any representation vector that subduces the identity of C$_{s}$ and that has
C$_{s}$ as its little group must have exactly one massless component. \ Now
consider the irrep spanned by $\{Z_{m}^{1^{-}}$\} where $Z_{m\pm }^{l}$ is a
tesseral harmonic. \ This subduces the group C$_{\infty v}$ once (which as
we have seen is its only little group). \ The subduction frequency of the
group C$_{nv}$ remains unity for all $n>1$. \ The group C$_{1v}$ is the
group C$_{s}$ and its identity representation is subduced twice by the irrep 
$Z_{m}^{1^{-}}$. \ However, according to the new rules, this is not a new
little group because the massive subduction frequency must be the ordinary
subduction frequency minus one (for the one massless component). \ Since the
massive subduction frequency remains equal to one, the shape of the
representation function and hence its little group cannot have changed.%
\textrm{\ }

However C$_{1}$ is not so clear. \ The basis functions of $\lambda =$ 1, the
tesseral harmonics $Z_{1+}^{l}$ and $Z_{1-}^{l}$, transform into each other
under rotations, and a linear combination of them has the same symmetry as
either of the functions individually. \ It follows that there is only one
little group for the $\ $irrep 1 and that its most general representation
vector is two-dimensional. \ The same argument applies to all irreps of this
group for $m>0$; a linear combination of the basis functions $\sin m\phi $
and $\cos m\phi $ is another sinusoidal function with the same period. \ The
little groups for each irrep, stated in brackets with the dimension of the
basis functions, are: C$_{\infty v}\left( 0,1\right) $; C$_{s}\left(
1,2\right) $; C$_{nv}\left( n,2\right) $.

A similar approach suffices in the cases of the two-dimensional irreps of D$%
_{\infty }$ and D$_{\infty h}$, and inspection of either basis function
gives each little group. For D$_{\infty }$ the little groups are D$_{\infty
}\left( A_{1},1\right) $, C$_{\infty }\left( A_{2},2\right) $, C$_{2}\left(
E_{1},2\right) $, D$_{n}\left( E_{n},2\right) .$For D$_{\infty h}$, we have D%
$_{\infty h}\left( A_{1}^{+},1\right) $, D$_{\infty }\left(
A_{1}^{-},1\right) $, C$_{\infty h}\left( A_{2}^{+},1\right) $, C$_{\infty
v}\left( A_{2}^{-},1\right) $; C$_{2h}\left( E_{1}^{\pm },2\right) $, D$%
_{nh}\left( E_{n}^{+},2\right) $ and D$_{nd}\left( E_{n}^{-},2\right) $ for $%
n$ even; D$_{nd}\left( E_{n}^{+},2\right) $ and D$_{nh}\left(
E_{n}^{-},2\right) $ for $n$ odd.

\section{Inspection of the O(3)\ little group analysis \label{Strategy}}

We claim the massive chain criterion to give the full solution of the
problem. However, we aim in this and the next section to illustrate and
confirm its conclusions by inspection of the results. Ultimately
establishing the sufficiency of $\ $a chain condition involves the explicit
demonstration of a function within the irrep space with the symmetry of the
candidate little group. This is a safeguard against producing yet another
abstract argument in favour of yet another inadequate criterion. We might
first inspect the symmetry elements of any function with respect to a global
axis choice, then convert the results of this inspection into a sufficiency
condition by showing that no rotation of axes will reveal new symmetry
elements.

In the following inspection-based approach is to the individual functions $%
\left\{ |\lambda l\rangle |l\in \lambda \right\} $ of the chosen basis for
the irrep space $\lambda $. Their maximal symmetries will be called the 
\emph{basis }little groups. For G = O(3), $\lambda =l^{\,\pi }$, and we
choose as basis the tesseral harmonics $|\lambda l\rangle \rightarrow \eta
Z_{m\pm }^{l\,}\left( \theta ,\phi \right) $ (suppressing the arguments for
simplicity), $m=0,1,2,..l$ with a scalar or pseudoscalar factor $\eta $ to
adjust the parity from polar to axial respectively (scalar if $\pi =\left(
-1\right) ^{l}$, pseudoscalar otherwise). Inspection of the symmetries and
identification of the little group of each such function is particularly
straightforward, the best choice of axes being conspicuous for such
fundamental functions. This nomenclature cannot avoid the arbitrariness of
the basis choice. However, an optimal or even near-optimal choice of basis
(in our case the tesseral harmonics) enables the identification of a maximal
number of little groups at the basis level, leaving fewer to be found in the
later steps.

We must consider other little groups corresponding to the symmetries of all
possible linear combinations $\sum_{l}a_{l}|\lambda l\rangle $ of the basis
vectors $\left\{ |\lambda l\rangle |l\in \lambda \right\} $. The literature
\ is ambivalent about the necessity of this, but we believe that any linear
combination is of as much interest in our applications as are the basis
functions. General linear combinations are those for which the same little
group applies to functions for a continuous range of the coefficients $%
\left\{ a_{l}\right\} $ of the basis functions $\left\{ |\lambda l\rangle
\right\} $. The consideration of general linear combinations is a trivial
problem for vectors; all linear combinations give another vector of the same
parity and so the same little group. Determining the symmetry of a general
linear combination is nontrivial, interesting and analytically solvable for
second rank tensors 2(SO(3)), i.e. the $l=2$ irrep of SO(3) (and similarly
for O(3)). Here the existence of higher symmetries for oblique axes is very
important. For higher irreps $l$ of O(3), we proceed by a combination of
inspection with guidance from the chain criterion and the group-chain
pedigree of any invariant function.

Suppose that two basis functions $|\lambda l_{1}\rangle $, $|\lambda
l_{2}\rangle $ with little groups H$_{1}$, H$_{2}$ are linearly combined.
The linear combination obviously includes all joint symmetries, and possibly
no others. Hence their common symmetry elements will make H$_{12}$ $=$ H$%
_{1}\cap $ H$_{2}$ a candidate little group. (The intersection $\cap $ needs
to be performed recognising that the axes of H$_{1}$ and H$_{2}$ may not be
parallel, this being already ascertained in the step of discovering the
basis little groups.) Hence the invariant count in H$_{12}$, a subgroup of H$%
_{1}$ and of H$_{2}$, satisfies $c_{\lambda }\left( \text{H}_{12}\right)
\geq c_{\lambda }\left( \text{H}_{1}\right) +c_{\lambda }\left( \text{H}%
_{2}\right) $. This relation follows from the linear independence of the
basis functions $|\lambda l_{1}\rangle $, $|\lambda l_{2}\rangle $ which
demonstrate the basis little group status of H$_{1}$ and H$_{2}$, because
each of these increments the invariant count. As little groups, H$_{1}$, H$%
_{2}$ have at least one invariant each, so that each term in this inequality
is nonzero. If for at least one of the group-subgroup paths H$_{1}\supset $ H%
$_{12}$, H$_{2}\supset $ H$_{12}$ any intermediate group H$^{\prime }$, if
it exists, has no invariants extra to those of the higher group H$_{1}$ or H$%
_{2}$, the invariant count $c_{\lambda }\left( \text{H}_{12}\right) $ is
greater than that of an adjacent higher group and the Michel chain criterion
is obeyed. However, before one can conclude that H$_{12}$ is a little group,
it needs to be verified that a linear combination of basis functions has no
further symmetries which are generated by the linear combination, and that
its symmetries are exhibited only by a suitable axis rotation.

As an example, consider linear combinations in $l=2$. $Z_{2-}^{2}\propto xy$
is a D$_{2}$ invariant in the chain O(3)-O$_{h}$-O-T-D$_{2}$ (forcing the C$%
_{2}^{\prime }$ axis through the cube edge) while $Z_{2+}^{2}\propto
x^{2}-y^{2}$ is a D$_{2}$ invariant in the chain O(3)-O$_{h}$-D$_{4h}$-D$%
_{4} $-D$_{2}$ (forcing the C$_{2}^{\prime }$ axis to a face centre). In
this way the possibilities for extra little groups become denumerable, as in
the chain criterion analysis, and can be given geometric character. Some of
the subtler points in implementing this are exemplified by Reid and Butler
(1982). The program \textsc{racah} was used to confirm the symmetry of some
more complicated basis functions. While this approach restricted as above to
a subset of axis choices and so still has a ``basis function'' flavour, the
axis choices are very much more numerous and apposite, and the results are
adequate in their variety to cover all the possible little groups revealed
by the chain criterion. For example, neither inspection of common symmetry
elements nor the program \textsc{racah} captures the retention of D$_{2}$
symmetry in a general linear combination of $Z_{2+}^{2}$ and $Z_{2-}^{2}$,
which is easily revealed by a suitable rotation about the $z$ axis; however
either of these functions suffices to illustrate the little group character
of D$_{2}$ in a way which is conveniently covered by its symmetry elements
under the standard axis choice.

Special linear combinations may exist in which only a unique set of
coefficients only will give the symmetry in question. The chain criterion of
Eq. \ref{Michel} is a reliable indicator since the requisite linear
combination will be unique to the group in question and the invariant count
will not be reflected in a supergroup. In the tesseral harmonic basis we
use, such special combinations may be expected at rank 4 (for a cubic
function necessarily involving the special linear combination of equation %
\ref{oh}).

C$_{i}$ and C$_{1}$ might be expected to be little groups in all irreps of
O(3), for positive and negative parity respectively, as the probable
symmetries (i.e. no rotational symmetry)\ of a fully general linear
combination of the basis functions. As noted below, this argument does not
work in the cases $l=1,2$. We now give a general approach to understanding
this, and the allied results for $l\geq 3$; and not only this, but a general
attack on the problem of dealing with the variation and uncertainty of the
axis choice that best reveals the symmetry of each linear combination.

If any linear combination $\sum_{m\pm }a_{m\pm }Z_{m\pm }^{l}$ is invariant
under a general rotation operator $O_{\phi }\left( \mathbf{n}\right) $ with
representation matrix $O_{mm^{\prime }}^{l}\left( \phi ,\mathbf{n}\right) $,
where $\phi $ has an arbitrary axis \textbf{n}, 
\begin{equation}
O_{mm^{\prime }}^{l}\left( \phi ,\mathbf{n}\right) a_{m^{\prime }}=a_{m}.
\label{eigen}
\end{equation}%
Hence $\left| \mathbf{O}^{l}\left( \phi ,\mathbf{n}\right) -\mathbf{I}%
\right| =0$, and because under a rotation of $\mathbf{n}$ to the $z$ axis, $%
\mathbf{O}^{l}\left( \phi ,\mathbf{n}\right) -\mathbf{I}=\mathbf{R}\left( 
\mathbf{O}^{l}\left( \phi ,\mathbf{z}\right) -\mathbf{I}\right) \mathbf{R}%
^{\dagger }$, the determinant of a matrix product is the product of the
determinants, $\left| \mathbf{O}^{l}\left( \phi ,\mathbf{z}\right) -\mathbf{I%
}\right| =0$. Conversely any rotation operator $O_{\phi }\left( \mathbf{z}%
\right) $ (and therefore any $O_{\phi }\left( \mathbf{n}\right) $)\
satisfies $\left| \mathbf{O}_{\phi }\left( \mathbf{n}\right) -\mathbf{I}%
\right| =0$ in the basis $Z_{m}^{l}$, since it leaves $Z_{0}^{l}$ unchanged.
We ask now the condition for a particular linear combination to be invariant
under any such rotation. A minimum condition is that both $\phi $ and $%
\mathbf{n}$ have to be chosen appropriately to the choice of $\left\{
a_{m}\right\} $. To explore this, for convenience we switch temporarily to a
spherical harmonic basis, in which $O_{mm^{\prime }}^{l}\left( \phi ,\mathbf{%
z}\right) =\delta _{mm^{\prime }}\exp \left( im\phi \right) $ and we use the
corresponding component labels $\pm m$. With this substitution equation (\ref%
{eigen}), while making no demand of $a_{0}$, requires the following
conditions on the possible linear combinations which are invariant under $%
O_{\phi }$. First $a_{1}=a_{-1}=0$; it is not possible for the factor $\exp
\left( \pm i\phi \right) -1$ to vanish for nontrivial $\phi $. This
vanishing of $a_{\pm 1}$ can always be achieved by a judicious choice of the
2 Euler angles (rotating the function $a_{m}Z_{m}^{l}$) and so defining the
direction of the axis $\mathbf{n}$. Since for $l=1$ this concludes the
requirements, and is always possible for any $\phi $ given an appropriate
axis $\mathbf{n}$, a general linear combination of $l=1$ functions always
has at least C$_{\infty }$ symmetry. This line of argument may be extended
to $l\geq 2$ (see Appendix 3), but the reliability of the massive chain
criterion (Section \ref{mass})\ reduces the importance of this to a check.

\section{Lower-order little groups of O(3) and SO(3)\label{o3}}

We distinguish polar (true)\ and axial tensors: polar tensors of rank $l$
reflect ``naturally,'' having parity $\pi =\pi _{l}\equiv \left( -1\right)
^{l}$, and axial tensors of that rank with parity $-\pi _{l}$. The basis
functions under consideration, with the relevant basis little groups as
established by inspection, are depicted in the various figures; little
groups for positive parity are on the left, and for negative parity on the
right, in each figure.

\subsection{$l=0,1$}

For $l=0$ we have trivially the basis function $\eta $ and the following
little groups.

\emph{Positive parity: 0}$^{+}$ \emph{(polar: invariant). }$\eta $ is a
scalar, and the little group is the full group O(3).

\emph{Negative parity; 0}$^{-}$\emph{\ (axial, pseudoscalar). }$\eta $ is a
pseudoscalar, and the little group is the full group SO(3).

For $l=1$, Figure 1 shows the angular dependence of the basis functions $%
\eta \left\{ Z_{m\pm }^{1}\right\} $ (the labels $m\pm $ denote the full set 
$0,1\pm ,...l\pm $ in general).

\emph{Positive parity 1}$^{+}$\emph{\ (pseudovector, axial). }$\eta $ is a
pseudoscalar.\emph{\ }By inspection the basis little group is C$_{\infty h}.$%
(This preserves an axial vector. Under $\sigma _{h}=\sigma _{xy},$ a
reflection in the $xy$ plane normal to the C$_{\infty }$ axis,\ a plain 1$%
^{-}$ spherical harmonic reverses; the pseudoscalar factor also reverses and
restores the sign. A magnetic field is unchanged by a reflection in a plane
to which it is perpendicular.)

\emph{Negative parity 1}$^{-}$\emph{\ (natural, polar). }$\eta $ is a
scalar. The basis little group is C$_{\infty v}$ which preserves a polar
vector in the reflection plane.

For either parity, any linear combination can only give another such vector,
with the same symmetry. Note that C$_{i}$ is not a little group of $1^{+}$,
or C$_{1}$ of 1$^{-}$; the symmetry of a general linear combination of $l=1$
functions is always at least C$_{\infty }$. This illustrates that the chain
criterion fails as a necessity condition, even for finite groups, since (in
the case of positive parity for example) the value (3) of $c_{1}\left( \text{%
C}_{i}\right) $ is greater than for any supergroup (Appendix 2). A proof
that C$_{\infty }$ is the minimum rotational symmetry is given in Section %
\ref{Strategy}.

We note that equations (\ref{mcc2}), (\ref{massless}) yield these results;
the $\dim V_{\lambda }-1$\emph{\ }restriction requires $f^{0}<3$. According
to the above $f^{0}=0$ for D$_{\infty h},$ C$_{\infty h},$ C$_{\infty v}$,
and according to Appendix 2\ the nonzero subduction frequencies are $c_{1+}($%
C$_{\infty h})=1=c_{1-}\left( \text{C}_{\infty v}\right) $\emph{. }

\subsection{$l=2$\label{o3-2}}

The case of $l=2$ is nontrivial. \ However, it is familiar in the form of
the representations found in the inertia tensor of rigid bodies in classical
mechanics (e.g. Goldstein, 1980 \S 5.4 or Landau \& Lifshitz 1976 vol. 1 \S
32). \ Such tensors contain both the 0 and 2 irreps of SO(3), and are
isomorphic to a real symmetric traceless 3$\times 3$ matrix. \ Figure 2
shows the basis functions $\eta \left\{ Z_{m\pm }^{2}\right\} $. The
functions $\eta \left\{ Z_{m\pm }^{2}|m>0\right\} $ are equal modulo a
rotation.

\emph{Positive parity 2}$^{+}$\emph{\ (natural, polar). }$\eta $ is a
scalar. The maximal symmetries of $\eta Z_{0}^{2}$ and $\eta \left\{ Z_{m\pm
}^{2}|m>0\right\} $ give D$_{\infty h}$ and D$_{2h}$ respectively. The axes
of the former are standard ($z$ for C$_{\infty }$, $x$ say for C$_{2}$ in D$%
_{\infty }$); the axes of the latter are oblique and $m$-dependent. This is
a dramatic example as to why merely working out intersections of groups is
not reliable in analysing linear combinations; the D$_{2}$ axes of the
functions $\left\{ Z_{m\pm }^{2}|m>0\right\} $ are all different. This also
illustrates why C$_{i}$ is not necessarily a little group of $l^{+}$.

\emph{Negative parity 2}$^{-}$\emph{\ (pseudotensor, axial). }$\eta $ is a
pseudoscalar. The maximal symmetries of $\eta Z_{0}^{2}$ and $\left\{ \eta
Z_{m\pm }^{2}|m>0\right\} $ are D$_{\infty }$ and D$_{2d}$ respectively. As
for the case of positive parity, the axes of the former are standard ($z$
for C$_{\infty }$, $x$ say for C$_{2}$ in D$_{\infty }$); the axes of the
latter are oblique and $m$-dependent.

The general linear combination of $\left\{ Z_{m\pm }^{2}\right\} $, $a\left(
2z^{2}-x^{2}-y^{2}\right) +$ $2bxy+2cyz+$ $2dzx+e\left( x^{2}-y^{2}\right) $
can be written $\left( x,y,z\right) M\left( x,y,z\right) ^{T}$ where the
matrix $M$ and its eignevector matrix $R$ are is

\begin{equation}
M=\left( 
\begin{array}{lll}
e-a & b & d \\ 
b & -e-a & c \\ 
d & c & 2a%
\end{array}%
\right) ,\,\,R=\left( 
\begin{array}{lll}
w_{11} & w_{21} & w_{31} \\ 
w_{12} & w_{22} & w_{32} \\ 
w_{13} & w_{23} & w_{33}%
\end{array}%
\right)
\end{equation}%
Since $M$ is symmetric and real (hermitian), its eigenvalues $\lambda _{a}$
are real and its eigenvectors $\mathbf{w}_{a}$ are orthogonal in the sense $%
\mathbf{w}_{a}^{\ast }\cdot \mathbf{w}_{b}=0$. Hence the eigenvector matrix $%
R$\ where $w_{ab}$ is the $b$th component of $w_{a}$ is unitary and also
diagonalises $M$:\ \ $RMR^{\dagger }=diag\left( \left\{ \lambda _{a}\right\}
\right) $. Since the matrix elements (as well as the the eigenvalues)\ are
real, the basis vectors can be chosen to be real, and $R$ is orthogonal and
so a rotation matrix. This defines a basis change under which such a linear
combination can be reduced to a linear combination of the diagonal terms $%
x^{2}$, $y^{2}$ and $z^{2}$, i.e. the functions $Z_{0}^{2}$, $Z_{2+}^{2}$,
whose combined rotational symmetry is D$_{2}.$ Hence D$_{2}$ is the minimal
symmetry of a general linear combination at $l=2$, and no new little groups
arise.

Another way of understanding this result is to note that symmetric functions
of second rank must be linear combinations of $x^{2},$ $xy$, $y^{2}$, $xz$, $%
yz$, $z^{2}$. Of these the invariant $x^{2}+y^{2}+z^{2}$ $\sim $ 0 can be
removed, leaving 5 rank two functions, according to the symmetric square of
the vector irrep in SO(3): 1$^{\times \left\{ 2\right\} }=2+0$. Three
coefficients (of $xy,yz,zx$ say)\ can be chosen to be zero by appropriate
choice of a rotation (the three Euler angles). This leaves two functions, $%
x^{2}-y^{2}\sim Z_{2+}^{2}$ and $3z^{2}-r^{2}\sim Z_{0}^{2}$.

Once again this illustrates the failure of the Michel chain criterion as a
necessity condition, since (in the case of positive parity for example) the
value (5) of $c_{2}\left( \text{C}_{i}\right) $ is greater than for any
supergroup (see Appendix 2). Another proof that D$_{2h}$ (for positive
parity)\ and D$_{2}$ (for negative parity)\ is the minimum symmetry is given
in Section \ref{Strategy}.

\subsection{$l=3$\label{l=3}}

It was proved in Section \ref{Strategy} that for $l\geq 3$, a truly general
linear combination cannot have any rotational symmetry element. Hence C$_{i}$
and C$_{1}$ are little groups for positive and negative parity respectively.

\emph{Positive parity 3}$^{+}$\emph{\ (pseudotensor, axial) }When we apply
the massive chain criterion (using the group branchings and subduction
frequencies of Appendix 2) for $l=3^{+}$, the only groups for which the
subduction frequency increases also exhibit an increase in the massive
subduction frequency. \ These groups are C$_{\infty h}$, T$_{h}$, D$_{3d}$, C%
$_{3i}$, C$_{2h}$ and C$_{i}$ .

These results are the same as those of Michel (1980 table A2) except that
they do not include the group D$_{2h}$. Ihrig and Golubitsky (1984) also
find the group D$_{2h}$ in this case. \ The reason that D$_{2h}$ is not
included here is that it is a subgroup of T$_{h}$, and the subduction
frequency is the same (1) for both these groups. \ This error probably arose
because the subduction frequency does increase in the chain D$_{\infty
h}\supset $D$_{nh}\supset $D$_{2h}$ because the subduction frequency for all
such supergroups of D$_{2h}$ is zero. \ This illustrates the need to test
the chain criterion for all possible supergroups.

\emph{Negative parity 3}$^{-}$\emph{\ (natural, polar)}

Application of the massive chain criterion to the case of the 7-dimensional
irrep of O(3) with negative parity yields the little groups: C$_{\infty v}$,
T$_{d}$, D$_{3h}$, C$_{3v},$ C$_{2v}$, C$_{s}$ and C$_{1}.$There are other
groups for which the subduction frequency increases. \ However in all of
these cases the massive subduction frequency is the same as that of a
supergroup. For example the group S$_{4}$ has a subduction frequency of 2,
but a massive subduction frequency of 1 in common with its supergroup T$_{d}$%
. \ Detailed inspection generates the same results as those above. \ The
detailed inspection method in 3$^{-}$ needs care over whether the group C$%
_{2}$ should be included in the set of little groups. \ Any linear
combination of $Z_{0}^{3}$, $Z_{2\pm }^{3}$ has C$_{2v}$ symmetry, because
this group is the intersection of their symmetry elements. \ But $Z_{3+}^{3}$
and $Z_{2-}^{3}$ have only C$_{2}$ for a common symmetry group, suggesting
that C$_{2}$ is a new little group for 3$^{-}.$ In fact an arbitrary linear
combination of these functions has the full C$_{2v}$ symmetry. \ Such a
linear combination can be generated from another that obviously has C$_{2v}$
symmetry, $a\left( Z_{0}^{3}+Z_{2}^{3}\right) +bZ_{-2}^{3}$, when rotated by
the transformation: $x\rightarrow -z,$ $y\rightarrow y,$ $z\rightarrow x$. \
This also shows that the choice of the two vectors with C$_{2h}$ symmetry
for 3$^{+}$ simply amounts to a choice of coordinate system. \ 

The basis functions connected with these little groups in 3$^{+},3^{-}$ are
listed in Appendix 3, because of their interest in calculation of property
tensors. Figure 3 gives the form of the full basis for $l=3$, and the tables
of Section \ref{gen} indicate the symmetries of these functions.

\subsection{$l=4$}

As for $l=3$, C$_{i}$ and C$_{1}$ are general little groups for positive and
negative parity respectively. We\thinspace inspect the symmetries of the
basis functions $\eta \left\{ Z_{m\pm }^{4}\right\} $ (Fig. 5, Table 3) for
the basis little groups.

\emph{Positive parity 4}$^{+}$. The basis little groups are D$_{\infty h}$, C%
$_{2h}$, D$_{2h}$, D$_{3d}$, D$_{4h}$ (for $\left| m\right| =0,1,2,3,4$
respectively).

\emph{Negative parity 4}$^{-}.$\ The basis little groups are D$_{\infty }$, C%
$_{2v}$, D$_{2d}$, D$_{3h}$, D$_{4d}$ (for $\left| m\right| =0,1,2,3,4$
respectively).

We form a histogram from the relevant column (4$\pm $) of Table 1 of the
numbers of invariants, and inspect the points of increase as the symmetry is
lowered. For\emph{\ positive }parity ($l^{\pi }=3^{+}$), D$_{\infty h}$, O$%
_{h}$ have 1 invariant. D$_{4h}$, D$_{3d}$ have\ 2 invariants. D$_{2h}$, C$%
_{3i}$, C$_{4h}$ have\ 3 invariants. C$_{2h}$ has 5 invariants. C$_{i}$ has\
7 invariants. This accounts for the basis and general little groups, and
also gives the candidates O$_{h}$, C$_{3i}$, C$_{4h}$.

O$_{h}$ is a little group, an example of such an invariant being the special
combination 
\begin{equation}
\zeta =\left( \sqrt{7}Z_{0}^{4}+\sqrt{5}Z_{4+}^{4}\right) /2\sqrt{3}
\label{oh}
\end{equation}%
Such results (see Appendix 3 for others)\ are most easily obtained from 
\textsc{racah}.

C$_{3i}$ possesses 3 invariants; it is a subgroup of 2 basis little groups, D%
$_{\infty h}$ and D$_{3d}$, giving two invariants from $Z_{0}^{4}$ and $%
Z_{3+}^{4}$. The third invariant therefore has to be sought elsewhere. Table
3 shows that the only possible additional function is $Z_{3-}^{4}$. However
its inclusion with $Z_{0}^{4}$ and $Z_{3+}^{4}$ can be frustrated by a
rotation about $z$, and the combination of $Z_{0}^{4}$ with either $%
Z_{3+}^{4}$ or $Z_{3-}^{4}$ has a higher symmetry. Hence C$_{3i}$ is not a
little group. (If we seek it from $\zeta $ of equation \ref{oh}, on the
grounds that C$_{3i}$ is also a subgroup of O$_{h}$, we then have to rotate $%
\zeta $ so that the C$_{3}$ axis is common, and the basis reduces to the set 
$Z_{0}^{4}$,$Z_{3+}^{4}$,$Z_{3-}^{4}$.)

This leaves C$_{4h}$. This is a subgroup of D$_{\infty h}$ and O$_{h}$
(whose invariants, involving as they do different linear combinations of $%
Z_{0}^{4}\ $and $Z_{4+}^{4}$, are linearly independent), and has a third
invariant. The 4-fold axis shows (from table 3)\ that this can only arise
from another linear combination of $Z_{0}^{4}\ $and $Z_{4\pm }^{4}$, but
this can be frustrated by a rotation about $z$. Hence C$_{4h}$ is not a
little group.

Hence O$_{h}$ is the only general\ little group for 4$^{+}$.

For \emph{negative parity }4$^{-}$(axial), D$_{\infty }$, O$,$ D$_{2d}$ and D%
$_{3h}$ have 1 invariant; D$_{4}$, D$_{3}$, S$_{4}$, C$_{3h}$ and C$_{2v}$
have\ 2 invariants; \emph{\ } D$_{2}$, C$_{3}$ and C$_{4}$ have\ 3
invariants; C$_{s}$ has\ 4 invariants; C$_{2}$ has 5 invariants, and C$_{1}$
has\ 7 invariants. This confirms the little group status of D$_{\infty }$, C$%
_{2v}$, D$_{2d}$, D$_{3h}$, D$_{4d}$ and C$_{1}$; this leaves O, D$_{4}$, D$%
_{3}$, D$_{2}$, S$_{4}$, C$_{3h}$, C$_{4}$, C$_{3}$, C$_{2}$, C$_{s}$ to be
discussed. We seek invariant functions for each of these groups from the set 
$\left\{ Z_{0}^{4},Z_{2\pm }^{4},Z_{3\pm }^{4},Z_{4\pm }^{4}\right\} $ as
before.

O is a little group from equation (\ref{oh}).

D$_{4}$ and C$_{4}$ both require a 4-fold axis, and so must be found in the
subset $\left\{ Z_{0}^{4},Z_{4\pm }^{4}\right\} $ A $z$ rotation eliminates
say $Z_{4-}^{4}$, and from Table 3 the remaining joint symmetry includes a
2-fold operation and so is indeed D$_{4}$, which becomes a little group.
Since this symmetry is higher than C$_{4}$, the latter is not a little
group; its candidacy assumed the necessity of admixing $Z_{4-}^{4}$. A
similar argument is true for S$_{4}$, the claim for whose candidacy also
requires both of $Z_{4\pm }^{4}$.

D$_{3}$, C$_{3h}$ and C$_{3}$ all require a 3-fold axis, and so must be
found in the subset $\left\{ Z_{0}^{4},Z_{3\pm }^{4}\right\} $ A $z$
rotation eliminates say $Z_{3-}^{4}$, and from Table 3 the remaining joint
symmetry elements, C$_{3z}$, C$_{2y}$ and C$_{2e}$ give D$_{3}$ as the only
little group. In particular, no reflection is possible. The appearance of C$%
_{3h}$ from the chain criterion is understandable on the basis of the
combination $Z_{3\pm }^{4}$, itself neutralised by a $z$ rotation, and
similarly C$_{3}$ on the basis of. combining $Z_{0}^{4}$, $Z_{3\pm }^{4}$.

The subset $\left\{ Z_{0}^{4},Z_{2+}^{4}\right\} $ similarly reveals D$_{2}$
as a little group, and the elimination of $Z_{2-}^{4}$ by rotation also
eliminates C$_{2}$.

This leaves C$_{s},$whose candidacy is explained by noting that the linear
combination of $\left\{ Z_{2-}^{4}\text{,}Z_{3-}^{4}\text{,}%
Z_{4-}^{4}\right\} $ has from Table 3 only the reflection symmetry $\sigma
_{xz}$. One of these three functions can be eliminated by $z$ rotation. If
we choose this to be $Z_{3-}^{4}$, not only is $\sigma _{xz}$ is common to
both the remaining functions, but C$_{2z}$ is also present. Hence C$_{s}$ is
not a little group.

Hence O, D$_{4}$, D$_{3}$ and D$_{2}$ are general\ little groups for 4$^{-}$.

\section{Little groups for any $l$\label{gen}}

The subgroups of O(3) and their group-subgroup relations are well known; in
seeking a general solution we require analytical formulae for the subduction
frequencies of these various groups. \ First, consider the tetrahedral group
T. \ From the Weyl trace formula from section 2.2 and noting from its
character table that T contains the following elements (with multiplicities
in brackets): C$_{1}$(1), C$_{3}$(4), C$_{3}^{2}$(4), C$_{2}$(3) the
subduction frequency is given by :

\begin{eqnarray*}
c_{\lambda }\left( \text{T}\right) &=&\tfrac{1}{12}\left( \frac{\sin \left(
\left( l+\frac{1}{2}\right) 2\pi \right) }{\sin \left( \pi \right) }+8\frac{%
\sin \left( \left( l+\frac{1}{2}\right) \frac{2\pi }{3}\right) }{\sin \left( 
\frac{\pi }{3}\right) }+3\frac{\sin \left( \left( l+\frac{1}{2}\right) \pi
\right) }{\sin \left( \frac{\pi }{2}\right) }\right) \\
&=&\tfrac{1}{12}\left( 2l+1\right) +\tfrac{4}{9}\sqrt{3}\left( \sin \frac{%
\pi }{3}\left( 2l+1\right) \right) +\tfrac{1}{4}\left( \sin \frac{\pi }{2}%
\left( 2l+1\right) \right) =2\left[ \frac{l}{3}\right] +\left[ \frac{l}{2}%
\right] -l+1.
\end{eqnarray*}%
where the last equation follows for $l$ being an integer. The square bracket
denotes the floor function, namely the largest integer less than or equal to
the argument. \ This result also applies to the group T$_{h}$ for positive
parity irreps.\ \ The same procedure allows us to determine formulae for
other subduction frequencies such as Y and O (also in SO(3); these hold also
for O(3)\ for either parity) and Y$_{h}$ and O$_{h}$ (for positive parity),
also a formula can be obtained for T$_{d}$ (for negative parity) as given in
Tables 1 and 2. For the moment we concentrate on SO(3) and Table 1.

Ihrig and Golubitsky (1984) augment the trace formula and correctly generate
subduction frequencies for groups such as C$_{n}$ for arbitrary values of $n$%
.\ For this presentation we proceed partly via inspection as deployed
initially by Ihrig and Golubitsky in the groups C$_{\infty }$ and C$_{n}$
and then apply the massive chain criterion. The dimension of the fixed point
set must be unity for C$_{\infty }$ because there is just one tesseral
harmonic basis function, $Z_{0}^{l},$ that is invariant under infinitesimal
rotations about an axis. \ From our earlier discussion the subduction
frequency for C$_{\infty }$ must be equal to the massive subduction
frequency (i.e. unity). \ Furthermore, as Ihrig and Golubitsky also observe,
the fixed point set for the group C$_{n}$ will include all basis functions
where $n$ divides $\left| m\right| $. \ It follows that the fixed point set
will always include the basis function $m=0$ and twice (one for $+m$ and one
for $-m$) the floor of $l$ divided by $n$. \ The case of C$_{1}$ has already
been dealt with; the conclusion being that it is a little group for all $%
l\geqslant 3$ with a subduction frequency of $2l+1$.

The subduction frequency of D$_{\infty }$ for all odd $l$ must equal zero as
only even $l$ basis functions contain even powers of $\cos \theta $ ($\theta 
$ being the polar variable) and so only they have the two-fold symmetry
perpendicular to the infinity fold axis found in the group D$_{\infty }$. \
Thus even $l$ irreps subduce the identity representation of D$_{\infty }$
once. \ The only other subgroups of SO(3) are the groups D$_{n}$ for all $%
n>1 $. \ If $l$ is even the fixed point set obviously includes $m=0$ (little
group D$_{\infty }$). \ It is clear from the case of C$_{n}$ that basis
functions for which $m$\ is an integer multiple of $n$ will contribute to
the fixed point set of D$_{n}$. \ However, they will only contribute to the
fixed point set in such a way that the two fold axes perpendicular to the $n$
fold axis coincide for all basis functions. \ This means that only one basis
function for a given $\left| m\right| $ will contribute to the fixed point
set. \ For even $l$ and even $n$ one may choose all appropriate $m$ to be
positive. \ For even $l$ and odd $n$ one may choose odd $m$ negative and
even $m$ positive. \ For odd $l$ one must exclude $m=0$ (little group C$%
_{\infty }$). \ If $n$ is even one may take all relevant $m$ to be negative.
\ When $n$ is odd one may choose the odd $m$ positive and the even $m$
negative. \ The preceding results with the massive chain criterion (Section %
\ref{mass}) produce the complete set of results for the little groups of
SO(3) in Table 1. The basis functions for the groups T, O and Y are not as
simply stated; only particular combinations of the tesseral harmonic basis
functions have the requisite symmetry. \ They may be determined for any
particular case by using the program \texttt{RACAH} (Appendix 3). In this
sense group branching arguments fully accommodate the effects of axis
freedom.

We now consider little groups of irreps of O(3). The results for positive
parity irreps are trivially derived from the little groups of SO(3) (see
caption to Table 2); the subduction frequencies and representation vectors
are exactly the same for these irreps. The case of negative parity may be
attacked in exactly the same way as that of positive parity. \ The only
caution applies to finding the general vectors for cases such as D$_{n}$
where one must again be sure that the two fold axes of different basis
functions coincide. \ Also, one must be sure that basis functions that are
included in the fixed point set do actually possess the requisite symmetry.
\ For example, the group D$_{4d}$ does not contain D$_{2d}$; the supergroups
of D$_{nd}$ are D$_{pnd}$ where $p$ is any odd positive integer. \ The cases
of D$_{nh}$, C$_{nh}$ and S$_{n}$ are similar. \ The results for $l>0$ are
given in Table 2. In our choice of basis we follow the convention (see
Section \ref{l=3}) in which the massless components that are eliminated are
those with $m=1\pm $ and one of the smallest $\left| m\right| >1$ possible
where necessary (often $m=2$-). \ 

The group O$_{h}$ is maximally connected to O(3); there are no subgroups of
O(3) that are supergroups of O$_{h}$. \ It follows that O$_{h}$ will be a
little group for any irrep that subduces its identity more that zero times
(except the identity irrep $l^{\pi }=0^{+}$). \ The same is true of the
group Y$_{h}$. \ In the case of the groups O, Y and T$_{d}$ one may note
that they are only connected to O(3) via inversion supergroups. \ It follows
that since these non-inversion groups can only be little groups of negative
parity irreps and since inversion groups are always subduced zero times by
negative parity irreps, that they will be little groups for any negative
parity irrep $l^{\pi }$, $l>0$ for which $c_{\lambda }>0.$ \ Since the
subduction frequency is partly a periodic function and partly a linear
function, the largest value of $l$ for which the subduction frequency could
be zero is obtained by comparing the linear part to the largest negative
value generated by the periodic part). \ This gives the little groups in the
first 5 rows of of Table 2.

The cases of T$_{h}$ for positive parity and T for negative parity require a
further observation. \ Suppose for some irrep of a group $G$ there exist at
least two little groups: $A$ and $B$. \ Further suppose that $A$ and $B$ are
not group--subgroup related. \ It follows that at least one of the massive
components of the most general representation vector with little group $A$
must be different from that with little group $B$. \ The representation
vector that subduces the identity representation of any subgroup of $A$ must
contain all of the components of the representation vector that subduces the
identity of $A$. \ The same is true for any subgroup of $B$. \ Now consider
a group $C$ that is a subgroup of both $A$ and $B$. \ It follows immediately
that the massive subduction frequency of $C$ must be strictly greater than
that of either $A$ or $B$ (whichever is the larger). \ Hence $C$ must also
be a little group of $G$ for the irrep under consideration. \ This means
that both T$_{h}$ (supergroups Y$_{h}$ and O$_{h}$) and T (supergroups Y, O
and T$_{d}$) must be little groups for all $l^{\pi }$ (respectively positive
and negative parity) greater than or equal to thirty. \ Since the subduction
frequencies given above must be the massive subduction frequencies for these
groups (which have non-collinear axes) it is a simple matter to apply the
massive chain criterion for all $l$ less than thirty. \ This procedure
determines that Table 2 gives the only irreps for which T$_{h}$ and T are
little groups:

\bigskip

{\Large Acknowledgements}

G\ E\ S is grateful to University of Canterbury for study leave and the
Department of Physics, University of Exeter for its hospitality during this
work. We are grateful to P\ H\ Butler and M\ F\ Reid for the use of \textsc{%
racah} (P\ H\ Butler and associates, 1995 \emph{The program} \textsc{racah}
v 3.1, Department of Physics and Astronomy, Christchurch, New Zealand).

\bigskip

{\LARGE Appendix 1: Comparisons with earlier work and chain criteria}

We give here some details of where previous work differs from the present
work, and a simple example of the use of subduction.

Boccara (1973) observes that the representation $l^{+}$ (for odd $l$)
subduces the identity of C$_{\infty }$. \ This is true, but C$_{\infty }$ is
not the relevant little group as claimed. \ Boccara states:
``Pseudo-tensorial order parameters cannot be used to characterise phases
with unproper symmetry elements.'' One counterexample is given by the
pseudotensorial irrep 1$^{+}$ of O(3) (the symmetry of the magnetic field)\
which contains the little group C$_{\infty h}$ and so an improper element
(the mirror reflection $\sigma _{h}$). \ Indeed, the irreducible Cartesian
tensor manifestation of all irreps $l^{+}$ will be pseudotensorial for all
odd $l$. \ If an irrep transforms as the identity (+) representation of the
inversion group, its little group must contain the inversion, an improper
symmetry element. \ Many of the pseudotensorial irreps that have negative
parity have little groups that contain reflection or rotation-reflection
elements. \ Boccara also neglects all of the octahedral, tetrahedral and
icosahedral groups, and we note that contrary to Boccara, 3$^{-}\downarrow
0\left( \text{D}_{2}\right) $.

Ihrig and Golubitsky (1984) use a group notation which is translated into
the Sch\"{o}nflies notation by: O$(2)\rightarrow $D$_{\infty },$ O$%
(2)^{-}\rightarrow $C$_{\infty \text{v}},$ I$\rightarrow $Y, O$%
^{-}\rightarrow $T$_{d},$ D$_{n}^{z}\rightarrow $C$_{nv},$ Z$_{n}\rightarrow 
$C$_{n}$, 1$\rightarrow $C$_{1}$; D$_{2n}^{d}\rightarrow $D$_{nd}$ when $n$
is even and D$_{2n}^{d}\rightarrow $D$_{nh}$ when $n$ is odd; D$%
_{2}^{d}\rightarrow C_{2v}$, Z$_{2n}^{-}\rightarrow $S$_{2n}$ when $n$ is
even and Z$_{2n}^{-}\rightarrow $C$_{nh}$ when $n$ is odd (Z$%
_{2}^{-}\rightarrow $C$_{s}$); the inversion group corresponding to the
rotation group G is $K\oplus $G. In SO(3), Ihrig and Golubitsky (1984) agree
with Michel (1980)\ except for the seven dimensional irrep: $l=3$. \ Here,
Ihrig erroneously\emph{\ }finds O where Michel (1980)\ correctly finds T. \
In common with Michel (1980)\ Ihrig and Golubitsky (1984)\ include
(erroneously, as we show)\ the group D$_{2}$ as a little group for $l=3$.
For rotation-reflection groups, a fundamental problem is that Ihrig and
Golubitsky do not specify the parity of the O(3) irreps under discussion. In
the case of postive parity, none of the little groups are correctly
identified. Entries in their Table 2 are inconsistent with their theorem 6.8
(Ihrig and Golubitsky 1984). For example, in their table 2 the groups listed
include D$_{3h}$ (for $l^{\pi }=3^{-}$) and D$_{5h}$ (for $l^{\pi }=5^{-}$).
\ Their theorem 6.8 (part f) includes the statement: ``D$_{2n}^{d} $, for $%
1<n<l$\ except D$_{4}^{d}$\ when $l=3$''. \ For D$_{3h}$ (for $l^{\pi
}=3^{-} $) and D$_{5h}$ (for $l^{\pi }=5^{-}$) to enter, this relation
should rather contain: $1<n\leqslant l$; the strict inequality would be an
error according to the present work, and is inconsistent with their
tabulation of C$_{2}$ and T\ under $l^{\pi }=3^{-}$. The present work agrees
with the first inequality as strict; however this is inconsistent with their
listing of D$_{2}^{d}$ for $l^{\pi }=5^{-}$. We retain the first inequality
as strict and interpret the strictness of their second inequality as a
typographical error. On this favourable reading, the little groups of O(3)
of theorem 6.8 of Ihrig and Golubitsky (1984) give results in agreement with
the present results for negative parity (only), except for their inclusion
of T as a little group of 3$^{-}$ and 4$^{-}.$ From Appendix 2, we note that
T$_{d}$ supplants T as the proper little group in 3$^{-}$, and O similarly
supplants T in 4$^{-}$; the relevant subduction frequencies do not increase
on descending from these groups to T. In each case, just this part of the
multiple branching diagram fails to satisfy the chain criterion.

Jaric (1986)\ quotes this work; his own work does not resolve these
problems. \emph{\ }Some minor corrections to Jaric (1986)\ may be noted.
Fig. 1 of Jaric covers all invariants constructible from powers of a rank $L$%
\ tensor via a sequence of ladder diagrams in which all lines have rank $L$.
Jaric's ladder construction gives an overdetermined set of invariants (in
fact an infinite set of relations), of which the first two alone are pure: $%
l=0,L$.\ The remaining ones are arbitrarily written linear combinations of
these and the other $2L-2$ independent invariants. There are linear
combinations between these when some or all the rank $L$\ terminals are
equal, corresponding to the reductions that occur when symmetrisation of the
product is made. If they are equal in pairs, the tree angular momenta are
even, halving the number of invariants from $2L$\ to $L.$\ When all four
terminals are equal, the further linear relations that must exist are
unknown; Jaric appeals to the integrity basis analysis result of Bistricky 
\emph{et al.} (1982) to show that the number of invariants is halved again.
It is better to have a single tree diagram as in Jaric's $\bar{I}_{L}^{4,1}$%
, not a sequence of ladder diagrams, and to allow all values $l=0,1,2$,..., $%
2L$\ of the coupled irreps consistent with Kronecker products, not just $L$.
This labels the possible independent invariants uniquely --- completely and
without repetition --- in the case that no symmetrisations are applicable
and the terminal tensors are different (though all of rank $L$). (Jaric's
eqs. 2.1 and 2.2 need a factor $2\sin ^{2}\phi /2$ in the integrand to give
the correct measure; this does not affect his conclusions.) Applications of
integrity basis theory are discussed by McLellan (1980), and graphical
methods have been used for integrity bases in SO($n$)\ by Ichinose and Ikeda
(1997). The integrity bases of O(3) have been extended to the case of
several tensors and related to angular momentum coupling trees for spin 1
and $\tfrac{1}{2}$ (Minard \emph{et al.} 1983, Riddell and Stedman 1984,
Stedman 1990).

Finally, we give a simple but nontrivial application of the chain criteria,
namely the little groups of the tetrahedral group T . \ Consider the group
chains T$\supset $D$_{2}\supset $C$_{2}\supset $C$_{1}$; T$\supset $ C$%
_{3}\supset $C$_{1}$. \ From character theory we obtain for irrep $A$ the
subduction frequencies $c_{A}($T, D$_{2}$, C$_{3}$, C$_{2}$ ,C$_{1})=1$.
Hence the little group of $A$ is T (as is appropriate for the identity
representation). \ For the irrep $E$ the frequencies are: $c_{E}($T$)=0$, $%
c_{E}($D$_{2})=1$, $c_{E}($C$_{2})=1$, $c_{E}($C$_{3})=0$, $c_{E}($C$_{1})=1$%
. Hence D$_{2}$ is the little group for $E$ (or $E^{^{\prime }}$, the
complex conjugate irrep). \ The 3-dimensional irrep $F$ has the subduction
frequencies: $c_{F}($T$)=0$, $c_{F}($D$_{2})=0$, $c_{F}($C$_{2})=1$, $c_{F}($%
C$_{3})=1$, $c_{F}($C$_{1})=3$. Hence one of the basis functions of $F$ has
the little group C$_{2}$ while another has the little group C$_{3}$. Finally
the linear combination of these two may be combined with the third basis
function, reducing the symmetry to that of the trivial little group C$_{1}$.
\ This gives the little groups of the irreps of the tetrahedral group as T$%
\left[ 1\right] $(A), D$_{2}\left[ 1\right] $(E,E$^{\prime }$), C$_{2}\left[
1\right] $, C$_{3}\left[ 1\right] $, C$_{1}\left[ 3\right] $(F). The numbers
in square brackets indicate the dimension of the most general vector in the
representation space of T that has this little group.

\bigskip

{\LARGE Appendix 2. Tables relevant to little group analysis}

Table 1 gives the subduction frequencies $c_{\lambda }\left( \text{H}\right) 
$ from the Weyl trace formula, as used for the traditional criteria. Tables
2 and 3 give the results of deducing the little groups from the massive
chain criterion.

\bigskip

{\LARGE Appendix 3. Inspection in higher irreps of SO(3), and basis functions%
}

If $l>1$, then in addition, either $a_{2}=a_{-2}=0$ or $\phi =\pm \pi $ (so
that $\exp \left( \pm 2i\phi \right) =1$), i.e. a 2-fold rotation only is
possible. Also for $l>1$, the freedom of choosing a rotation by 3 Euler
angles leaves one degree of freedom, the third Euler angle, to be exploited
as required. In the case $l=2$ we can choose the third Euler angle to make
the coefficient of say $Z_{-2}^{l}$ zero; this leaves the functions $%
Z_{0}^{2}$, $Z_{2+}^{2}$ as adequate to describe any linear combination.
Since any linear combination of these functions has the D$_{2}$ rotational
symmetry which they share, this is the minimum rotational symmetry in $l=2$.
In addition these functions (with even $m$) have compatible reflection
symmetries, so that in the case of even parity a general linear combination
has D$_{2h}$ symmetry.

If $l>2$, then in addition, either $a_{3}=a_{-3}=0$ or $\phi =\pm 2\pi /3$,
i.e. only a 3-fold rotation can evade the necessity of requiring $a_{\pm
3}=0 $. Since a 2-fold and 3-fold rotation are not compatible, it is
impossible to avoid imposing at least 3 conditions on $a_{\pm 2}$ and $%
a_{\pm 3}$ (3 rather than 4, because of the possibility of using the third
Euler angle to cancel the fourth); and so it is impossible for a general
linear combination at $l=3$ to have a nontrivial rotational symmetry.
However, the symmetry of a general linear combination of the full basis set $%
\left\{ Z_{m\pm }^{3}\right\} $ can be simplified by an appropriate axis
choice to restrict the number of functions under consideration to $\left\{
Z_{0}^{3},Z_{2\pm }^{3},Z_{3\pm }^{3}\right\} $ with the free choice of one
linear relation between the coefficients $a_{2\pm }$, $a_{3\pm }$ (for
example $a_{2-}=0$). Since the reflection symmetries of $Z_{3\pm }^{3}$
(where $m$ is odd)\ are incompatible with those of the other functions (with
even $m$), no reflection plane can survive in general in the case $l>2$.
Hence C$_{i}$ and C$_{1}$ are the maximal symmetries of a general linear
form and and so little groups for positive and negative parity respectively
for $l>2$.

This proceeds similarly for higher $l$; for $l=4$ we can consider a general
linear combination as a combination of $\left\{ Z_{0}^{4},Z_{2\pm
}^{4},Z_{3\pm }^{4},Z_{4\pm }^{4}\right\} $ with the coefficient of one of
the $Z_{m-}^{4}$ functions being set to zero. Special functions and their
little groups at $l=4$ include equation (\ref{oh}) for O\ and at $l=6$
include Y$:\left( \sqrt{11}Z_{0}^{6}-\sqrt{14}Z_{5}^{6}\right) /5;$ O$%
:\left( -Z_{0}^{6}+\sqrt{7}Z_{4}^{6}\right) /\sqrt{8};$ T$:\left( -Z_{0}^{6}+%
\sqrt{7}Z_{4}^{6}\right) /\sqrt{8},$ $\left( -\sqrt{11}Z_{2}^{6}+\sqrt{5}%
Z_{6}^{6}\right) /4.$

Since the little groups of positive parity irreps are determined purely by
their rotational symmetry we need only consider pure rotations in using the
inspection method. \ The tesseral harmonic basis functions in 3$^{+}$ have
the following symmetries (confirmed by the program \texttt{RACAH}): $%
Z_{0}^{3}$: C$_{\infty h}$, $Z_{1\pm }^{3}$: C$_{2h}$, $Z_{2\pm }^{3}$: T$%
_{h}$, $Z_{3\pm }^{3}$: D$_{3d}$. A general linear combination of $%
Z_{m+}^{l} $ with $Z_{m-}^{l}$ will always have the same symmetry as $%
Z_{m+}^{l}$ for the reasons discussed in Section \ref{Strategy}. \ We may
now consider all possible linear combinations of the above. \ This is
essentially a combinatorial exercise although we use the fact that it is
always possible to remove $Z_{1\pm }^{3}\ $ and one other function with $m>0$%
, say $Z_{m-}^{3}$\ by performing the appropriate rotation. \ The results
are as follows: \ $Z_{0}^{3},Z_{2+}^{3}$: C$_{2h},$ $Z_{0}^{3},Z_{3+}^{3}$: C%
$_{3i}, $ $Z_{2-}^{3},Z_{3+}^{3}$: C$_{2h},$ $Z_{2+}^{3},Z_{3\pm }^{3}$: C$%
_{i},$ $Z_{0}^{3},Z_{2+}^{3}Z_{3\pm }^{3}$: C$_{i}$. \ In short, the massive
chain criterion finds all possible little groups. \ The massive subduction
frequency also gives the number of massive components of the most general
representation vector that has a particular little group. \ The vectors
which are appropriate for different $l$, H are indicated in Tables 1,2. The
convention by which the massless components are removed is that of Section %
\ref{l=3}, \ref{gen}. According to the above we could have chosen $%
Z_{-2}^{3},Z_{3}^{3}$ as the vector for C$_{2h}$ instead of $%
Z_{0}^{3},Z_{2}^{3}$. \ These two vectors actually belong to the same
stratum.

\newpage

{\Large References}

\noindent%
Andrews D\ L, Naguleswaran S and Stedman G E 1998 Phys. Rev. A \textbf{57}
4925-4929.

\noindent%
Birman J\ 1966 Phys. Rev. Lett. \textbf{17} 1216-1219

\noindent%
---\ 1982 Physica A \textbf{114} 564-571

\noindent%
Boccara N 1973 Ann. Phys. \textbf{76} 72-79.

\noindent%
Churcher C\ D and Stedman G\ E\ 1981 \emph{\ }J. Phys. C \textbf{14}
2237--2264

\noindent%
Constantinescu D H 1979 J de Physique \textbf{40} 147-159

\noindent%
Cracknell A\ P, Lorenc J and Przystawa J\ A 1976 J. Phys. C \textbf{9}
1731-1758

\noindent%
Fraleigh J\ B 1994 \emph{A first course in abstract algebra} (5th ed.)
(Addison-Wesley, Reading, Mass.)

\noindent%
Gaeta G 1984 Phys. Rev. B \textbf{29}\ 6371

\noindent%
Giordmaine J\ A 1965 Phys. Rev. \textbf{138A} 1599-1606

\noindent%
Girardi G, Sciarrino A and Sorba P 1982 Physica A \textbf{114} 370-388

\noindent%
Goldrich F\ E\ and Birman J\ 1968 Phys. Rev. \textbf{167} 528-532

\noindent%
Ichinose S and Ikeda N 1997 J. Math. Phys. \textbf{38} 6457-6521

\noindent%
Ihrig E\ and Golubitsky M 1984 Physica D \textbf{13} 1-33

\noindent%
Jacobi CC\ G\ J\ 1834 Poggendorf Ann. Phys. Chem. \textbf{33}\ 229-238

\noindent%
Jaric M\ V\ 1982 Physica A \textbf{114} 550-556

\noindent%
---\ 1983a J. Math. Phys. \textbf{24} 2865-2882

\noindent%
--- 1983b Phys. Rev. Lett, \textbf{51} 2073-2076

\noindent%
--- 1986 Nucl. Phys. B \textbf{26} 647-670

\noindent%
Jaric M\ V and Senechal M 1984 J. Math. Phys. \textbf{25} 3148-3154

\noindent%
Keller J\ B\ and Antman S 1969\emph{\ Bifurcation theory and nonlinear
eigenvalue problems }(Benjamin:\ New York)

\noindent%
Kim Y\ S\ 1996 Acta Phys. Polon. B \textbf{27 }2741-2746

\noindent%
--- 1997 Int. J. Mod. Phys. A \textbf{12 }71-78

\noindent%
Konig A and Mermin N D 1997 Phys. Rev. B \textbf{56 }13607-13610

\noindent%
Landsman N\ P\ and Wiedemann U\ A 1995 Rev. Math. Phys. \textbf{7} 923-958

\noindent%
Leaf B 1998 Found. Phys. \textbf{11 }1-22

\noindent%
Lorenc J, Przystawa J\ A and Cracknell A\ P 1980 \textbf{13 }1955-1961

\noindent%
Loudon R\ 1964 Adv. Phys. \textbf{13 }423-482

\noindent%
McLellan A\ G\ 1980 \emph{The classical thermodynamics of deformable
materials }(Cambridge Univ. Press:\ Cambridge)

\noindent%
Michel L 1980 Rev. Mod. Phys. \textbf{52} 640-651

\noindent%
Minard R\ A, Stedman G\ E\ and McLellan A G 1983 J. Chem. Phys. \textbf{78}
5016--5024

\noindent%
Nash P L 1997 Nuovo Cim. B \textbf{112 }1415-1421

\noindent%
Newman D\ J\ 1971 Adv. Phys. \textbf{20} 197-256

\noindent%
Nye J\ F\ 1957 \emph{Physical properties of crystals} (Clarendon: Oxford)

\noindent%
Poincare H\ 1885 Acta Math. \textbf{7}\ 259

\noindent%
Przystawa J\ 1982 Physica A \textbf{114} 557-563

\noindent%
Reid\ M\ F and Butler PH\ 1982 J Phys A \textbf{15} 2327-35

\noindent%
Rembielinski J 1997 Int. J. Mod. Phys. A \textbf{12 }1677-1709

\noindent%
Riddell A\ G\ and Stedman G E 1984 Phys. Rev. A \textbf{30} 1727--1733

\noindent%
Sattinger D\ H 1978 J Math. Phys. \textbf{19} 1720-1732

\noindent%
Stedman G\ E\ 1985 \emph{\ }Adv. Phys. \textbf{34} 513--587

\noindent%
---\ 1990 \emph{Diagram techniques in group theory} (Cambridge Univ. Press:
Cambridge)

\noindent%
Vollhardt D and W\"{o}lfle P 1990 \emph{The superfluid phases of Helium 3}
(Taylor and Francis:\ London)

\noindent%
Yariv A and Yeh P\ 1984 \emph{Optical waves in crystals} (Wiley:\ New\ York)

\newpage

\section{Tables}

\bigskip

Table 1. Irreps of SO(3) for which the stated groups are little groups when $%
l>0$. The special cases for Y are $l=$ 6, 10, 12, 15, 16, 18, 20-22, 24-28.
Representation vectors for Y, O, T must be determined from \textsc{racah}
for each $l$ (see Appendix 3 for examples). $\kappa \equiv \left( -1\right)
^{k}$.

$\bigskip $

\[
\begin{tabular}{lllll}
H & $c_{l}\left( \text{H}\right) $ & General & Special $l$ & Representation
vectors \\ \hline\hline
Y & $\left[ \frac{l}{5}\right] +\left[ \frac{l}{3}\right] +\left[ \frac{l}{2}%
\right] -l+1$ & $l\geqslant 30$ & caption & - \\ 
O & $\left[ \frac{l}{4}\right] +\left[ \frac{l}{3}\right] +\left[ \frac{l}{2}%
\right] -l+1$ & $l\geqslant 12$ & 4, 6, 8, 9, 10 & - \\ 
T & $2\left[ \frac{l}{3}\right] +\left[ \frac{l}{2}\right] -l+1$ & $%
l\geqslant 9$ & 3, 6, 7 & - \\ 
D$_{\infty }$ & 1 & $l$ even & - & $Z_{0}^{l}$ \\ 
C$_{\infty }$ & 1 & $l$ odd & - & $Z_{0}^{l}$ \\ 
D$_{n}$ & $\left[ \frac{l}{n}\right] $ & $l$ odd, 2 $\leqslant n\leqslant
l\geqslant 4$ & $l=n=3$ & $Z_{n,-\kappa }^{l},Z_{2n,-\kappa
}^{l},Z_{3n,-\kappa }^{l}$... \\ 
D$_{n}$ & $\left[ \frac{l}{n}\right] +1$ & $l$ even, 2 $\leqslant n\leqslant
l\geqslant 4$ & $l=n=2$ & $Z_{0}^{l},Z_{n\kappa }^{l},Z_{2n\kappa
}^{l},Z_{3n\kappa }^{l}$... \\ 
C$_{n}$ & $2\left[ \frac{l}{n}\right] +1$ & $l$ even, $2$ $\leqslant
n\leqslant l/2$ & - & $Z_{0}^{l},Z_{n+}^{l},Z_{2n+}^{l},Z_{3n+}^{l}$... \\ 
C$_{n}$ & $2\left[ \frac{l}{n}\right] +1$ & $l$ odd, $2$ $\leqslant
n\leqslant l$ & - & $Z_{0}^{l},Z_{n+}^{l},Z_{2n+}^{l},Z_{3n+}^{l}$... \\ 
C$_{1}$ & $2l+1$ & $l\geqslant 3$ & - & $Z_{0}^{l},Z_{2+}^{l},Z_{3\pm
}^{l},Z_{4\pm }^{l},...$%
\end{tabular}%
\]

\bigskip

Table 2. Irreps $l^{-}$ (O(3)) for which the stated (non-inversion)\ groups
are little groups when $l>0$. The little groups for irreps $l^{+}$(O(3))\
are the inversion groups, and the results may be obtained from Table 1 with
the mappings D$_{\infty }\rightarrow $D$_{\infty h},$ C$_{\infty
}\rightarrow $C$_{\infty h}$, Y$\rightarrow $Y$_{h}$, O$\rightarrow $O$_{h},$
T$\rightarrow $T$_{h}$, (D$_{n}\rightarrow $D$_{nh}$, C$_{n}\rightarrow $C$%
_{nh}$ if $n$ is even), (D$_{n}\rightarrow $D$_{nd}$, C$_{n}\rightarrow $C$%
_{ni}$ if $n$ is odd). The special cases for Y are $l=$ 6, 10, 12, 15, 16,
18, 20-22, 24-28. $,\kappa \equiv \left( -1\right) ^{k}$, $\mu \equiv \left(
-1\right) ^{l}$, $\nu \equiv \kappa \mu $.

$\bigskip $

\[
\begin{tabular}{lllll}
H & $c_{l^{\pi }}\left( \text{H}\right) $ & General & Special cases & 
Representation vectors \\ \hline\hline
Y & $\left[ \frac{l}{5}\right] +\left[ \frac{l}{3}\right] +\left[ \frac{l}{2}%
\right] -l+1$ & $l\geqslant 30$ & caption & - \\ 
O & $\left[ \frac{l}{4}\right] +\left[ \frac{l}{3}\right] +\left[ \frac{l}{2}%
\right] -l+1$ & $l\geqslant 12$ & 4, 6, 8, 9, 10 & - \\ 
T$_{d}$ & $\left[ \frac{l+2}{4}\right] +\left[ \frac{l}{3}\right] +\left[ 
\frac{l+1}{2}\right] -l$ & $l\geqslant 9$ & 3, 6, 7 & - \\ 
T & $2\left[ \frac{l}{3}\right] +\left[ \frac{l}{2}\right] -l+1$ & $%
l\geqslant 12$ & 6, 9, 10 & - \\ 
D$_{\infty }$ & 1 & $l$ even & - & $Z_{0}^{l}$ \\ 
C$_{\infty v}$ & 1 & $l$ odd & - & $Z_{0}^{l}$ \\ 
D$_{nd}$ & $\left[ \left( l+n\right) /2n\right] $ & $n$ even, 2 $\leqslant
n\leqslant l\geqslant 4$ & $n=l=2$ & $Z_{n\nu }^{l},Z_{3n\nu }^{l},Z_{5n\nu
}^{l}$... \\ 
D$_{nh}$ & $\left[ \left( l+n\right) /2n\right] $ & $n$ odd, 2 $\leqslant
n\leqslant l\geqslant 4$ & $n=l=3$ & $Z_{n\nu }^{l},Z_{3n\nu }^{l},Z_{5n\nu
}^{l}$... \\ 
D$_{n}$ & $\left[ \frac{l}{n}\right] +1$ & $l$ even, 2 $\leqslant n\leqslant
l$ & - & $Z_{0}^{l},Z_{n\kappa }^{l},Z_{2n\kappa }^{l},Z_{3n\kappa }^{l}$...
\\ 
D$_{n}$ & $\left[ \frac{l}{n}\right] $ & $l$ odd, 2 $\leqslant n\leqslant
l/2 $ & - & $Z_{n,-\kappa }^{l},Z_{2n,-\kappa }^{l},Z_{3n,-\kappa }^{l}$...
\\ 
C$_{nv}$ & $\left[ \frac{l}{n}\right] $ & $l$ even, 2 $\leqslant n\leqslant
l/2$ & - & $Z_{n-}^{l},Z_{2n-}^{l},Z_{3n-}^{l}$... \\ 
C$_{nv}$ & $\left[ \frac{l}{n}\right] +1$ & $l$ odd, 2 $\leqslant n\leqslant
l$ & - & $Z_{0}^{l},Z_{n+}^{l},Z_{2n+}^{l},Z_{3n+}^{l}$... \\ 
C$_{nh}$ & $2\left[ \frac{l+n}{2n}\right] $ & $n$ odd, $n\leqslant l/3$ & -
& $Z_{n+}^{l},Z_{3n\pm }^{l},Z_{5n\pm }^{l}...$ \\ 
C$_{s}$ & $2\left[ \frac{l+n}{2n}\right] $ & $l$ even, $l\geqslant 4$ & - & $%
Z_{0}^{l},Z_{2+}^{l},Z_{3+}^{l},...Z_{l+}^{l}$ \\ 
C$_{s}$ & $2\left[ \frac{l+n}{2n}\right] $ & $l$ odd, $l\geqslant 3$ & - & $%
Z_{2-}^{l},Z_{3-}^{l},...Z_{l-}^{l}$ \\ 
S$_{2n}$ & $2\left[ \frac{l+n}{2n}\right] $ & $2\leqslant n\leqslant l/3$ & -
& $Z_{n+}^{l},Z_{3n\pm }^{l},Z_{5n\pm }^{l}...$ \\ 
C$_{n}$ & $2\left[ \frac{l}{n}\right] +1$ & $2\leqslant n\leqslant l/2$ & -
& $Z_{0}^{l},Z_{n\pm }^{l},Z_{2n\pm }^{l},...$ \\ 
C$_{1}$ & $2l+1$ & $l\geqslant 3$ & - & $Z_{0}^{l},Z_{2+}^{l},Z_{3\pm }^{l},$%
...,$Z_{l\pm }^{l}$%
\end{tabular}%
\]

\newpage

Table 3. Adjacent group branchings H$\supset $K, H$\subset $H$^{\prime }$,
and (except for the column headed $\bar{f}_{\lambda }$) subduction
frequencies $c_{\lambda }\left( \text{H}\right) $ for the irreps $l^{\pi }$,
within O(3) for $l\leqslant 6$. \ These frequencies were calculated from the
computer program \texttt{RACAH} v3.1 (Butler 1995). The corresponding
massive subduction frequency $f_{\lambda }^{m}\left( \text{H}\right) $\ is
obtained by subtracting the massless subduction frequency $f_{\lambda
}^{0}\left( \text{H}\right) $ from the tabulated entries (equation \ref%
{massdef}). From equation (\ref{massless}) $f_{\lambda }^{0}\left( \text{H}%
\right) $ is the smaller of $2l$ and of $\bar{f}_{\lambda }\left( \text{H}%
\right) $, which is also tabulated.

\hspace{-2.5cm}%
\[
\begin{tabular}{llllllllllllllll}
{\small H} & {\small Adjacent groups}\qquad \qquad\ $\ $ & $\bar{f}_{\lambda
}$ & {\small 0}$^{-}$ & {\small 1}$^{+}$ & {\small 1}$^{-}$ & {\small 2}$%
^{+} $ & {\small 2}$^{-}$ & {\small 3}$^{+}$ & {\small 3}$^{-}$ & {\small 4}$%
^{+}$ & {\small 4}$^{-}$ & {\small 5}$^{+}$ & {\small 5}$^{-}$ & {\small 6}$%
^{+}$ & {\small 6}$^{-}$ \\ \hline\hline
{\small C}$_{1}$ & $\subset ${\small C}$_{5}${\small , C}$_{2}${\small , C}$%
_{i}${\small , C}$_{s},${\small \ C}$_{3}$ & {\small 3} & {\small 1} & 
{\small 3} & {\small 3} & {\small 5} & {\small 5} & {\small 7} & {\small 7}
& {\small 9} & {\small 9} & {\small 11} & {\small 11} & {\small 13} & 
{\small 13} \\ 
{\small C}$_{i}$ & $\supset ${\small C}$_{1}${\small ; }$\subset ${\small C}$%
_{5i}${\small , C}$_{2h}${\small , C}$_{3i}$ & {\small 3} & {\small 0} & 
{\small 3} & {\small 0} & {\small 5} & {\small 0} & {\small 7} & {\small 0}
& {\small 9} & {\small 0} & {\small 11} & {\small 0} & {\small 13} & {\small %
0} \\ 
{\small C}$_{s}$ & $\supset ${\small C}$_{1}${\small ; }$\subset ${\small C}$%
_{5h}${\small ,C}$_{5v}${\small ,C}$_{2h}${\small ,C}$_{2v}${\small ,C}$%
_{3v} $ & {\small 1} & {\small 0} & {\small 1} & {\small 2} & {\small 3} & 
{\small 2} & {\small 3} & {\small 4} & {\small 5} & {\small 4} & {\small 5}
& {\small 6} & {\small 7} & {\small 6} \\ 
{\small C}$_{2}$ & $\supset ${\small C}$_{1}${\small ; }$\subset ${\small C}$%
_{4}${\small , C}$_{2h}${\small , D}$_{3}${\small , D}$_{2}${\small , C}$%
_{6} $ & {\small 1} & {\small 1} & {\small 1} & {\small 1} & {\small 3} & 
{\small 3} & {\small 3} & {\small 3} & {\small 5} & {\small 5} & {\small 5}
& {\small 5} & {\small 7} & {\small 7} \\ 
{\small C}$_{2h}$ & $\supset ${\small C}$_{2}${\small ,C}$_{i}${\small ,C}$%
_{s}${\small ;}$\subset ${\small C}$_{4h}${\small ,D}$_{2h}${\small ,C}$%
_{6h} ${\small ,D}$_{3d}$ & {\small 1} & {\small 0} & {\small 1} & {\small 0}
& {\small 3} & {\small 0} & {\small 3} & {\small 0} & {\small 5} & {\small 0}
& {\small 5} & {\small 0} & {\small 7} & {\small 0} \\ 
{\small C}$_{2v}$ & $\supset ${\small C}$_{2}${\small , C}$_{s}${\small ; }$%
\subset ${\small C}$_{4v}${\small ,D}$_{2d}${\small ,D}$_{2h}${\small ,C}$%
_{6v}$ & {\small 0} & {\small 0} & {\small 0} & {\small 1} & {\small 2} & 
{\small 1} & {\small 1} & {\small 2} & {\small 3} & {\small 2} & {\small 2}
& {\small 3} & {\small 4} & {\small 3} \\ 
{\small C}$_{3}$ & $\supset ${\small C}$_{1}${\small ; }$\subset ${\small T,D%
}$_{3}${\small ,C}$_{6}${\small ,C}$_{3i}${\small ,C}$_{3v}${\small ,C}$%
_{3h} $ & {\small 1} & {\small 1} & {\small 1} & {\small 1} & {\small 1} & 
{\small 1} & {\small 3} & {\small 3} & {\small 3} & {\small 3} & {\small 3}
& {\small 3} & {\small 5} & {\small 5} \\ 
{\small C}$_{3h}$ & $\supset ${\small C}$_{3}${\small , C}$_{s}${\small ; }$%
\subset ${\small D}$_{3h}${\small , C}$_{6h}$ & {\small 1} & {\small 0} & 
{\small 1} & {\small 0} & {\small 1} & {\small 0} & {\small 1} & {\small 2}
& {\small 1} & {\small 2} & {\small 1} & {\small 2} & {\small 3} & {\small 2}
\\ 
{\small C}$_{3i}$ & $\supset ${\small C}$_{3}${\small , C}$_{i}${\small ; }$%
\subset ${\small T}$_{h}${\small , C}$_{6h}${\small , D}$_{3d}$ & {\small 1}
& {\small 0} & {\small 1} & {\small 0} & {\small 1} & {\small 0} & {\small 3}
& {\small 0} & {\small 3} & {\small 0} & {\small 3} & {\small 0} & {\small 5}
& {\small 0} \\ 
{\small C}$_{3v}$ & $\supset ${\small C}$_{3}${\small ,C}$_{s}${\small ;}$%
\subset ${\small T}$_{d}${\small ,C}$_{6v}${\small ,D}$_{3h}${\small ,D}$%
_{3d}$ & {\small 0} & {\small 0} & {\small 0} & {\small 1} & {\small 1} & 
{\small 0} & {\small 1} & {\small 2} & {\small 2} & {\small 1} & {\small 1}
& {\small 2} & {\small 3} & {\small 2} \\ 
{\small C}$_{4}$ & $\supset ${\small C}$_{2}${\small ; }$\subset ${\small C}$%
_{\infty }${\small , C}$_{4h}${\small , C}$_{4v}${\small , D}$_{4}$ & 
{\small 1} & {\small 1} & {\small 1} & {\small 1} & {\small 1} & {\small 1}
& {\small 1} & {\small 1} & {\small 3} & {\small 3} & {\small 3} & {\small 3}
& {\small 3} & {\small 3} \\ 
{\small C}$_{4h}$ & $\supset ${\small C}$_{4}${\small , C}$_{2h}${\small , S}%
$_{4}${\small ; }$\subset ${\small C}$_{\infty h}${\small , D}$_{4h}$ & 
{\small 1} & {\small 0} & {\small 1} & {\small 0} & {\small 1} & {\small 0}
& {\small 1} & {\small 0} & {\small 3} & {\small 0} & {\small 3} & {\small 0}
& {\small 3} & {\small 0} \\ 
{\small C}$_{4v}$ & $\supset ${\small C}$_{4}${\small , C}$_{2v}${\small ; }$%
\subset ${\small C}$_{\infty v}${\small , D}$_{4d}${\small , D}$_{4h}$ & 
{\small 0} & {\small 0} & {\small 0} & {\small 1} & {\small 1} & {\small 0}
& {\small 0} & {\small 1} & {\small 2} & {\small 1} & {\small 1} & {\small 2}
& {\small 2} & {\small 1} \\ 
{\small C}$_{5}$ & $\supset ${\small C}$_{1}${\small ;}$\subset ${\small C}$%
_{5i}${\small ,C}$_{5h}${\small ,C}$_{5v}${\small ,D}$_{5}${\small ,C}$%
_{\infty }$ & {\small 1} & {\small 1} & {\small 1} & {\small 1} & {\small 1}
& {\small 1} & {\small 1} & {\small 1} & {\small 1} & {\small 1} & {\small 3}
& {\small 3} & {\small 3} & {\small 3} \\ 
{\small C}$_{5h}$ & $\supset ${\small C}$_{5}${\small , C}$_{s}${\small ; }$%
\subset ${\small D}$_{5h}$ & {\small 1} & {\small 0} & {\small 0} & {\small 1%
} & {\small 1} & {\small 0} & {\small 1} & {\small 0} & {\small 1} & {\small %
0} & {\small 1} & {\small 2} & {\small 1} & {\small 2} \\ 
{\small C}$_{5i}$ & $\supset ${\small C}$_{5}${\small , C}$_{i}${\small ; }$%
\subset ${\small D}$_{5d}${\small , C}$_{\infty h}$ & {\small 1} & {\small 0}
& {\small 1} & {\small 0} & {\small 1} & {\small 0} & {\small 1} & {\small 0}
& {\small 1} & {\small 0} & {\small 3} & {\small 0} & {\small 3} & {\small 0}
\\ 
{\small C}$_{5v}$ & $\supset ${\small C}$_{5}${\small , C}$_{s}${\small ; }$%
\subset ${\small C}$_{\infty v}${\small , D}$_{5d}$ & {\small 0} & {\small 0}
& {\small 0} & {\small 1} & {\small 1} & {\small 0} & {\small 0} & {\small 1}
& {\small 1} & {\small 0} & {\small 1} & {\small 2} & {\small 2} & {\small 1}
\\ 
{\small C}$_{6}$ & $\supset ${\small C}$_{3}${\small , C}$_{2}${\small ; }$%
\subset ${\small C}$_{\infty }${\small , D}$_{6}${\small , C}$_{6h}${\small %
, C}$_{6v}$ & {\small 1} & {\small 1} & {\small 1} & {\small 1} & {\small 1}
& {\small 1} & {\small 1} & {\small 1} & {\small 1} & {\small 1} & {\small 1}
& {\small 1} & {\small 3} & {\small 3} \\ 
{\small C}$_{6h}$ & $\supset ${\small C}$_{6}${\small ,C}$_{3h}${\small ,C}$%
_{3i}${\small ,C}$_{2h}${\small ;}$\subset ${\small C}$_{\infty h}${\small ,D%
}$_{6h}$ & {\small 1} & {\small 0} & {\small 1} & {\small 0} & {\small 1} & 
{\small 0} & {\small 1} & {\small 0} & {\small 1} & {\small 0} & {\small 1}
& {\small 0} & {\small 3} & {\small 0} \\ 
{\small C}$_{6v}$ & $\supset ${\small C}$_{6}${\small , C}$_{2v}${\small , C}%
$_{3v}${\small ; }$\subset ${\small C}$_{\infty v}${\small , D}$_{6h}$ & 
{\small 0} & {\small 0} & {\small 0} & {\small 1} & {\small 1} & {\small 0}
& {\small 0} & {\small 1} & {\small 1} & {\small 0} & {\small 0} & {\small 1}
& {\small 2} & {\small 1} \\ 
{\small C}$_{\infty }$ & $\supset ${\small C}$_{6}${\small , C}$_{5}${\small %
, C}$_{4}${\small ; }$\subset ${\small C}$_{\infty h}${\small , D}$_{\infty
} $ & {\small 0} & {\small 1} & {\small 1} & {\small 1} & {\small 1} & 
{\small 1} & {\small 1} & {\small 1} & {\small 1} & {\small 1} & {\small 1}
& {\small 1} & {\small 1} & {\small 1} \\ 
{\small C}$_{\infty h}$ & $\supset ${\small C}$_{\infty }${\small , C}$_{6h}$%
{\small , C}$_{5i}${\small , C}$_{4h}${\small ; }$\subset ${\small D}$%
_{\infty h}$ & {\small 0} & {\small 0} & {\small 1} & {\small 0} & {\small 1}
& {\small 0} & {\small 1} & {\small 0} & {\small 1} & {\small 0} & {\small 1}
& {\small 0} & {\small 1} & {\small 0} \\ 
{\small C}$_{\infty v}$ & $\supset $ {\small C}$_{\infty }$, {\small C}$%
_{6v} ${\small , C}$_{5v}${\small , C}$_{4v}${\small ; }$\subset ${\small D}$%
_{\infty h}$ & {\small 0} & {\small 0} & {\small 0} & {\small 1} & {\small 1}
& {\small 0} & {\small 0} & {\small 1} & {\small 1} & {\small 0} & {\small 0}
& {\small 1} & {\small 1} & {\small 0}%
\end{tabular}%
\]

\bigskip

\hspace{-2.5cm}%
\[
\begin{tabular}{llllllllllllllll}
{\small H} & {\small Adjacent groups}\qquad \qquad $\ \ \ \ \ $ & $\bar{f}%
_{\lambda }$ & {\small 0}$^{-}$ & {\small 1}$^{+}$ & {\small 1}$^{-}$ & 
{\small 2}$^{+}$ & {\small 2}$^{-}$ & {\small 3}$^{+}$ & {\small 3}$^{-}$ & 
{\small 4}$^{+}$ & {\small 4}$^{-}$ & {\small 5}$^{+}$ & {\small 5}$^{-}$ & 
{\small 6}$^{+}$ & {\small 6}$^{-}$ \\ \hline\hline
{\small D}$_{2}$ & $\supset ${\small C}$_{2}${\small ; }$\subset ${\small D}$%
_{2d}${\small , T, D}$_{4}${\small , D}$_{2h}${\small , D}$_{6}$ & {\small 0}
& {\small 1} & {\small 0} & {\small 0} & {\small 2} & {\small 2} & {\small 1}
& {\small 1} & {\small 3} & {\small 3} & {\small 2} & {\small 2} & {\small 4}
& {\small 4} \\ 
{\small D}$_{2d}$ & $\supset ${\small D}$_{2}${\small , C}$_{2v}${\small , S}%
$_{4}${\small ; }$\subset ${\small D}$_{4h}${\small , D}$_{6d}${\small , T}$%
_{d}$ & {\small 0} & {\small 0} & {\small 0} & {\small 0} & {\small 1} & 
{\small 1} & {\small 0} & {\small 1} & {\small 2} & {\small 1} & {\small 1}
& {\small 1} & {\small 2} & {\small 2} \\ 
{\small D}$_{2h}$ & $\supset ${\small D}$_{2}${\small , C}$_{2h}${\small , C}%
$_{2v}${\small ; }$\subset ${\small D}$_{4h}${\small , D}$_{6h}${\small , T}$%
_{h}$ & {\small 0} & {\small 0} & {\small 0} & {\small 0} & {\small 2} & 
{\small 0} & {\small 1} & {\small 0} & {\small 3} & {\small 0} & {\small 2}
& {\small 0} & {\small 4} & {\small 0} \\ 
{\small D}$_{3}$ & $\supset ${\small C}$_{3}${\small , C}$_{2}${\small ; }$%
\subset ${\small O, Y, D}$_{6}${\small , D}$_{3d}${\small , D}$_{3h}$ & 
{\small 0} & {\small 1} & {\small 0} & {\small 0} & {\small 1} & {\small 1}
& {\small 1} & {\small 1} & {\small 2} & {\small 2} & {\small 1} & {\small 1}
& {\small 3} & {\small 3} \\ 
{\small D}$_{3d}$ & $\supset ${\small D}$_{3}${\small ,C}$_{3i}${\small ,C}$%
_{3v}${\small ,C}$_{2h}${\small ;}$\subset ${\small O}$_{h}${\small ,Y}$_{h}$%
{\small ,D}$_{6h}$ & {\small 0} & {\small 0} & {\small 0} & {\small 0} & 
{\small 1} & {\small 0} & {\small 1} & {\small 0} & {\small 2} & {\small 0}
& {\small 1} & {\small 0} & {\small 3} & {\small 0} \\ 
{\small D}$_{3h}$ & $\supset ${\small D}$_{3}${\small , C}$_{3h}${\small , C}%
$_{3v}${\small ; }$\subset ${\small D}$_{6h}${\small , D}$_{6d}$ & {\small 0}
& {\small 0} & {\small 0} & {\small 0} & {\small 1} & {\small 0} & {\small 0}
& {\small 1} & {\small 1} & {\small 1} & {\small 0} & {\small 1} & {\small 2}
& {\small 1} \\ 
{\small D}$_{4}$ & $\supset ${\small D}$_{2}${\small , C}$_{4}${\small ; }$%
\subset ${\small D}$_{\infty }${\small , O, D}$_{4h}$ & {\small 0} & {\small %
1} & {\small 0} & {\small 0} & {\small 1} & {\small 1} & {\small 0} & 
{\small 0} & {\small 2} & {\small 2} & {\small 1} & {\small 1} & {\small 2}
& {\small 2} \\ 
{\small D}$_{4d}$ & $\supset ${\small D}$_{4}${\small , C}$_{4v}${\small ; }$%
\subset ${\small D}$_{\infty h}$ & {\small 0} & {\small 0} & {\small 0} & 
{\small 0} & {\small 1} & {\small 0} & {\small 0} & {\small 0} & {\small 1}
& {\small 1} & {\small 0} & {\small 1} & {\small 1} & {\small 1} \\ 
{\small D}$_{4h}$ & $\supset ${\small D}$_{4}${\small ,C}$_{4h}${\small ,C}$%
_{4v}${\small ,D}$_{2h}${\small ,D}$_{2d}${\small ;}$\subset ${\small O}$%
_{h} ${\small ,D}$_{\infty h}$ & {\small 0} & {\small 0} & {\small 0} & 
{\small 0} & {\small 1} & {\small 0} & {\small 0} & {\small 0} & {\small 2}
& {\small 0} & {\small 1} & {\small 0} & {\small 2} & {\small 0} \\ 
{\small D}$_{5}$ & $\supset ${\small C}$_{5}${\small , C}$_{2}${\small ; }$%
\subset ${\small D}$_{5d}${\small , D}$_{5h}${\small , D}$_{\infty }${\small %
, Y} & {\small 0} & {\small 1} & {\small 0} & {\small 0} & {\small 1} & 
{\small 1} & {\small 0} & {\small 0} & {\small 1} & {\small 1} & {\small 1}
& {\small 1} & {\small 2} & {\small 2} \\ 
{\small D}$_{5d}$ & $\supset ${\small D}$_{5}${\small , C}$_{5i}${\small , C}%
$_{5v}${\small , C}$_{2h}${\small ; }$\subset ${\small D}$_{\infty h}$ & 
{\small 0} & {\small 0} & {\small 0} & {\small 0} & {\small 1} & {\small 0}
& {\small 0} & {\small 0} & {\small 1} & {\small 0} & {\small 1} & {\small 0}
& {\small 2} & {\small 0} \\ 
{\small D}$_{5h}$ & $\supset ${\small D}$_{5}${\small , C}$_{5h}${\small ; }$%
\subset ${\small D}$_{\infty h}$ & {\small 0} & {\small 0} & {\small 0} & 
{\small 0} & {\small 1} & {\small 0} & {\small 0} & {\small 0} & {\small 1}
& {\small 0} & {\small 0} & {\small 1} & {\small 1} & {\small 1} \\ 
{\small D}$_{6}$ & $\supset ${\small D}$_{3}${\small , D}$_{2}${\small , C}$%
_{6}${\small ; }$\subset ${\small D}$_{\infty }${\small , D}$_{6h}${\small ,
D}$_{6d}$ & {\small 0} & {\small 1} & {\small 0} & {\small 0} & {\small 1} & 
{\small 1} & {\small 0} & {\small 0} & {\small 1} & {\small 1} & {\small 0}
& {\small 0} & {\small 2} & {\small 2} \\ 
{\small D}$_{6d}$ & $\supset ${\small D}$_{6}${\small , D}$_{2d}${\small ; }$%
\subset ${\small D}$_{\infty h}$ & {\small 0} & {\small 0} & {\small 0} & 
{\small 0} & {\small 1} & {\small 0} & {\small 0} & {\small 0} & {\small 1}
& {\small 0} & {\small 0} & {\small 0} & {\small 1} & {\small 1} \\ 
{\small D}$_{6h}$ & $\supset ${\small D}$_{6}${\small ,D}$_{2h}${\small ,D}$%
_{3h}${\small ,D}$_{3d}${\small ,C}$_{6h}${\small ,C}$_{6v}${\small ;}$%
\subset ${\small D}$_{\infty h}$ & {\small 0} & {\small 0} & {\small 0} & 
{\small 0} & {\small 1} & {\small 0} & {\small 0} & {\small 0} & {\small 1}
& {\small 0} & {\small 0} & {\small 0} & {\small 2} & {\small 0} \\ 
{\small D}$_{\infty }$ & $\supset ${\small C}$_{\infty }${\small , D}$_{6}$%
{\small , D}$_{5}${\small , D}$_{4}${\small ; }$\subset ${\small D}$_{\infty
h}${\small , SO(3)} & {\small 0} & {\small 1} & {\small 0} & {\small 0} & 
{\small 1} & {\small 1} & {\small 0} & {\small 0} & {\small 1} & {\small 1}
& {\small 0} & {\small 0} & {\small 1} & {\small 1} \\ 
{\small D}$_{\infty h}$ & $\supset ${\small D}$_{\infty }${\small ,C}$%
_{\infty v}${\small ,D}$_{6h}${\small ,D}$_{5d}${\small ,D}$_{4h}${\small ;}$%
\subset ${\small O(3)} & {\small 0} & {\small 0} & {\small 0} & {\small 0} & 
{\small 1} & {\small 0} & {\small 0} & {\small 0} & {\small 1} & {\small 0}
& {\small 0} & {\small 0} & {\small 1} & {\small 0} \\ 
{\small Y} & $\supset ${\small T, D}$_{5}${\small , D}$_{3}${\small , C}$%
_{5v}${\small ; }$\subset ${\small SO(3), Y}$_{h}$ & {\small 0} & {\small 1}
& {\small 0} & {\small 0} & {\small 0} & {\small 0} & {\small 0} & {\small 0}
& {\small 0} & {\small 0} & {\small 0} & {\small 0} & {\small 1} & {\small 1}
\\ 
{\small Y}$_{h}$ & $\supset ${\small Y, D}$_{5d}${\small ; }$\subset $%
{\small O(3)} & {\small 0} & {\small 0} & {\small 0} & {\small 0} & {\small 0%
} & {\small 0} & {\small 0} & {\small 0} & {\small 0} & {\small 0} & {\small %
0} & {\small 0} & {\small 1} & {\small 0} \\ 
{\small O} & $\supset ${\small T, D}$_{4}${\small , D}$_{3}${\small ; }$%
\subset ${\small SO(3), O}$_{h}$ & {\small 0} & {\small 1} & {\small 0} & 
{\small 0} & {\small 0} & {\small 0} & {\small 0} & {\small 0} & {\small 1}
& {\small 1} & {\small 0} & {\small 0} & {\small 1} & {\small 1} \\ 
{\small O(3)} & $\supset ${\small SO(3), D}$_{\infty h}${\small , Y}$_{h}$%
{\small , O}$_{h}$ & {\small 0} & {\small 0} & {\small 0} & {\small 0} & 
{\small 0} & {\small 0} & {\small 0} & {\small 0} & {\small 0} & {\small 0}
& {\small 0} & {\small 0} & {\small 0} & {\small 0} \\ 
{\small O}$_{h}$ & $\supset ${\small O,T}$_{d}${\small ,T}$_{h}${\small ,D}$%
_{3d}${\small ,D}$_{4h}${\small ,C}$_{3i}${\small ;}$\subset ${\small O(3)}
& {\small 0} & {\small 0} & {\small 0} & {\small 0} & {\small 0} & {\small 0}
& {\small 0} & {\small 0} & {\small 1} & {\small 0} & {\small 0} & {\small 0}
& {\small 1} & {\small 0} \\ 
{\small S}$_{4}$ & $\supset ${\small C}$_{2}${\small ; }$\subset ${\small C}$%
_{4h}${\small , D}$_{2d}$ & {\small 1} & {\small 0} & {\small 1} & {\small 0}
& {\small 1} & {\small 2} & {\small 1} & {\small 2} & {\small 3} & {\small 2}
& {\small 3} & {\small 2} & {\small 3} & {\small 4} \\ 
{\small SO(3)} & $\supset ${\small D}$_{\infty }${\small , Y, O; }$\subset $%
{\small O(3)} & {\small 0} & {\small 1} & {\small 0} & {\small 0} & {\small 0%
} & {\small 0} & {\small 0} & {\small 0} & {\small 0} & {\small 0} & {\small %
0} & {\small 0} & {\small 0} & {\small 0} \\ 
{\small T} & $\supset ${\small D}$_{2}${\small , C}$_{3}${\small ; }$\subset 
${\small O, Y, T}$_{h}${\small , T}$_{d}$ & {\small 0} & {\small 1} & 
{\small 0} & {\small 0} & {\small 0} & {\small 0} & {\small 1} & {\small 1}
& {\small 1} & {\small 1} & {\small 0} & {\small 0} & {\small 2} & {\small 2}
\\ 
{\small T}$_{d}$ & $\supset ${\small T, D}$_{2d}${\small ; }$\subset $%
{\small O}$_{h}$ & {\small 0} & {\small 0} & {\small 0} & {\small 0} & 
{\small 0} & {\small 0} & {\small 0} & {\small 1} & {\small 1} & {\small 0}
& {\small 0} & {\small 0} & {\small 1} & {\small 1} \\ 
{\small T}$_{h}$ & $\supset ${\small T, D}$_{2h}${\small , C}$_{3i}${\small %
; }$\subset ${\small O}$_{h}${\small , Y}$_{h}$ & {\small 0} & {\small 0} & 
{\small 0} & {\small 0} & {\small 0} & {\small 0} & {\small 1} & {\small 0}
& {\small 1} & {\small 0} & {\small 0} & {\small 0} & {\small 2} & {\small 0}%
\end{tabular}%
\]

\bigskip

\newpage Table 4. Little groups H of irreps of SO(3) for $l\leq 4$, and for $%
n\leq 4$ in D$_{n}$, C$_{n}$; a prescription for general $l,n$ is in Table
1. Nonzero entries indicate little groups, the numbers being the appropriate
subduction frequencies $c_{\lambda }$.

\bigskip 
\[
\begin{tabular}{llllll}
$l$ & 0 & 1 & 2 & 3 & 4 \\ \hline\hline
SO(3) & 1 & - & - & - & - \\ 
D$_{\infty }$ & - & - & 3 & - & 3 \\ 
C$_{\infty }$ & - & 3 & - & 3 & - \\ 
O & - & - & - & - & 4 \\ 
T & - & - & - & 4 & - \\ 
D$_{4}$ & - & - & - & - & 5 \\ 
D$_{3}$ & - & - & - & 4 & 5 \\ 
D$_{2}$ & - & - & 5 & - & 6 \\ 
C$_{3}$ & - & - & - & 5 & - \\ 
C$_{2}$ & - & - & - & 5 & 7 \\ 
C$_{1}$ & - & - & - & 7 & 9%
\end{tabular}%
\]

\bigskip

\newpage Table 5. Little groups H of irreps of O$\left( 3\right) $ for $%
l\leq 9$, and for $n\leq 9$ in D$_{n}$, C$_{n}$; a prescription for general $%
l,n$ is in Table 2. Nonzero entries indicate little groups, the numbers
being the appropriate subduction frequencies.$c_{\lambda }.$

\begin{quote}
\bigskip

\[
\begin{tabular}{cccccccccccc}
H & $l^{+}$ & 0 & 1 & 2 & 3 & 4 & 5 & 6 & 7 & 8 & 9 \\ \hline\hline
{O(3)} &  & {1} & {-} & {-} & {-} & {-} & {-} & {-} & {-} & {-} & - \\ 
{D$_{\infty h}$} &  & {-} & {-} & {3} & {-} & {3} & {-} & {3} & {-} & {3} & -
\\ 
{C$_{\infty h}$} &  & {-} & {3} & {-} & {3} & {-} & {3} & {-} & {3} & {-} & 3
\\ 
{Y$_{h}$} &  & {-} & {-} & {-} & {-} & {-} & {-} & {4} & {-} & {-} & - \\ 
{O$_{h}$} &  & {-} & {-} & {-} & {-} & {4} & {-} & {4} & {-} & {4} & 4 \\ 
{T$_{h}$} &  & {-} & {-} & {-} & {4} & {-} & {-} & {5} & {4} & {-} & 5 \\ 
{D$_{9d}$} &  & {- } & {-} & {-} & {-} & {-} & {-} & {-} & {-} & {-} & 4 \\ 
{D$_{8h}$} &  & {- } & {-} & {-} & {-} & {-} & {-} & {-} & {-} & {5} & 4 \\ 
{D$_{7d}$} &  & {- } & {-} & {-} & {-} & {-} & {-} & {-} & {4} & {5} & 4 \\ 
{D$_{6h}$} &  & {- } & {-} & {-} & {-} & {-} & {-} & {5} & {4} & {5} & 4 \\ 
{D$_{5d}$} &  & {- } & {-} & {-} & {-} & {-} & {4} & {5} & {4} & {5} & 4 \\ 
{D$_{4h}$} &  & {- } & {-} & {-} & {-} & {5} & {4} & {5} & {4} & {6} & 5%
\end{tabular}%
\]

\bigskip

\[
\ 
\begin{tabular}{cccccccccccc}
H & $l^{+}$ & 0 & 1 & 2 & 3 & 4 & 5 & 6 & 7 & 8 & 9 \\ \hline\hline
{D$_{3d}$} &  & {- } & {-} & {-} & {4} & {5} & {4} & {6} & {5} & {6} & 6 \\ 
{D$_{2h}$} &  & {- } & {-} & {5} & {-} & {6} & {5} & {7} & {6} & {8} & 7 \\ 
{C$_{9i}$} &  & {- } & {-} & {-} & {-} & {-} & {-} & {-} & {-} & {-} & 5 \\ 
{C$_{8h}$} &  & {- } & {-} & {-} & {-} & {-} & {-} & {-} & {-} & {-} & 5 \\ 
{C$_{7i}$} &  & {- } & {-} & {-} & {-} & {-} & {-} & {-} & {5} & {-} & 5 \\ 
{C$_{6h}$} &  & {- } & {-} & {-} & {-} & {-} & {-} & {-} & {5} & {-} & 5 \\ 
{C$_{5i}$} &  & {- } & {-} & {-} & {-} & {-} & {5} & {-} & {5} & {-} & 5 \\ 
{C$_{4h}$} &  & {- } & {-} & {-} & {-} & {-} & {5} & {-} & {5} & {7} & 7 \\ 
{C$_{3i}$} &  & {- } & {-} & {-} & {5} & {-} & {5} & {7} & {7} & {7} & 9 \\ 
{C$_{2h}$} &  & {- } & {-} & {-} & {5} & {7} & {7} & {9} & {9} & {11} & 11
\\ 
{C$_{i}$} &  & {-} & {-} & {-} & {7} & {9} & {11} & {13} & {15} & {17} & 19%
\end{tabular}%
\]

\bigskip

\[
\begin{tabular}{cccccccccccc}
H & $l^{-}$ & 0 & 1 & 2 & 3 & 4 & 5 & 6 & 7 & 8 & 9 \\ \hline\hline
SO(3) &  & 1 & - & - & - & - & - & - & - & - & - \\ 
D$_{\infty }$ &  & - & - & 3 & - & 3 & - & 3 & - & 3 & - \\ 
C$_{\infty v}$ &  & - & 3 & - & 3 & - & 3 & - & 3 & - & 3 \\ 
Y &  & - & - & - & - & - & - & 4 & - & - & - \\ 
O &  & - & - & - & - & 4 & - & 4 & - & 4 & 4 \\ 
T$_{d}$ &  & - & - & - & 4 & - & - & 4 & 4 & - & 4 \\ 
T &  & - & - & - & - & - & - & 5 & - & - & 5 \\ 
D$_{9h}$ &  & - & - & - & - & - & - & - & - & - & 4 \\ 
D$_{9}$ &  & - & - & - & - & - & - & - & - & - & - \\ 
D$_{8d}$ &  & - & - & - & - & - & - & - & - & 4 & 4 \\ 
D$_{8}$ &  & - & - & - & - & - & - & - & - & 5 & - \\ 
D$_{7h}$ &  & - & - & - & - & - & - & - & 4 & 4 & 4 \\ 
D$_{7}$ &  & - & - & - & - & - & - & - & - & 5 & - \\ 
D$_{6d}$ &  & - & - & - & - & - & - & 4 & 4 & 4 & 4 \\ 
D$_{6}$ &  & - & - & - & - & - & - & 5 & - & 5 & - \\ 
D$_{5h}$ &  & - & - & - & - & - & 4 & 4 & 4 & 4 & 4 \\ 
D$_{5}$ &  & - & - & - & - & - & - & 5 & - & 5 & - \\ 
D$_{4d}$ &  & - & - & - & - & 4 & 4 & 4 & 4 & 4 & 4 \\ 
D$_{4}$ &  & - & - & - & - & 5 & - & 5 & - & 5 & 5%
\end{tabular}%
\ 
\]%
\newline
\[
\begin{tabular}{cccccccccccc}
H & $l^{-}$ & 0 & 1 & 2 & 3 & 4 & 5 & 6 & 7 & 8 & 9 \\ \hline\hline
{D$_{3h}$} &  & {- } & {-} & {-} & {4} & {4} & {4} & {4} & {4} & {4} & 5 \\ 
{D$_{3}$} &  & {-} & {-} & {-} & {-} & {5} & {-} & {5} & {5} & {5} & 6 \\ 
{D$_{2d}$} &  & {- } & {-} & {4} & {-} & {4} & {4} & {5} & {5} & {5} & 5 \\ 
{D$_{2}$} &  & {-} & {-} & {5} & {-} & {5} & {5} & {6} & {6} & {6} & 7 \\ 
{C$_{9v}$} &  & {- } & {-} & {-} & {-} & {-} & {-} & {-} & {-} & {-} & 5 \\ 
{C$_{8v}$} &  & {- } & {-} & {-} & {-} & {-} & {-} & {-} & {-} & {-} & 5 \\ 
{C$_{7v}$} &  & {- } & {-} & {-} & {-} & {-} & {-} & {-} & {5} & {-} & 5 \\ 
{C$_{6v}$} &  & {- } & {-} & {-} & {-} & {-} & {-} & {-} & {5} & {-} & 5 \\ 
{C$_{5v}$} &  & {- } & {-} & {-} & {-} & {-} & {5} & {-} & {5} & {-} & 5 \\ 
{C$_{4v}$} &  & {- } & {-} & {-} & {-} & {-} & {5} & {-} & {5} & {5} & 6 \\ 
{C$_{3v}$} &  & {- } & {-} & {-} & {5} & {-} & {5} & {5} & {6} & {5} & 7 \\ 
{C$_{2v}$} &  & {- } & {-} & {-} & {5} & {5} & {6} & {6} & {7} & {7} & 8 \\ 
{S$_{4}$} &  & {-} & {-} & {-} & {-} & {-} & {-} & {6} & {6} & {6} & 6 \\ 
{C$_{3h}$} &  & {- } & {-} & {-} & {-} & {-} & {-} & {-} & {-} & {-} & 6 \\ 
{C$_{s}$} &  & {-} & {-} & {-} & {6} & {6} & {8} & {8} & {10} & {10} & 12 \\ 
{C$_{4}$} &  & {-} & {-} & {-} & {-} & {-} & {-} & {-} & {-} & {7} & 7 \\ 
{C$_{3}$} &  & {-} & {-} & {-} & {-} & {-} & {-} & {7} & {7} & {7} & 9 \\ 
{C$_{2}$} &  & {-} & {-} & {-} & {-} & {7} & {7} & {9} & {9} & {11} & 11 \\ 
{C$_{1}$} &  & {-} & {-} & {-} & {7} & {9} & {11} & {13} & {15} & {17} & 19%
\end{tabular}%
\]
\end{quote}

\bigskip

\bigskip \newpage

\bigskip

\textbf{Captions to figures}

Figure 1. Basis functions of $l=0,$1 (O(3)). Some of the relevant little
groups are indicated below, the bracketed indices indicating the changes
between positive and negative parity in O(3) and between O(3) and SO(3).

Figure 2. Basis functions of $l=2$ (O(3)) as for figure 1.

Figure 3. Basis functions of $l=3$ (O(3)) as for figure 1.

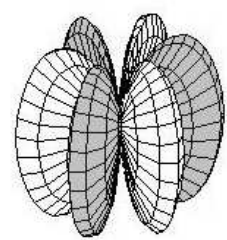
\end{document}